\RequirePackage{fix-cm}
\documentclass[fleqn,usenatbib]{mnras}
\usepackage{newtxtext,newtxmath}

\usepackage[T1]{fontenc}

\DeclareRobustCommand{\VAN}[3]{#2}
\let\VANthebibliography\thebibliography
\def\thebibliography{\DeclareRobustCommand{\VAN}[3]{##3}\VANthebibliography}

\usepackage{graphicx}	
\usepackage{amsmath}	

\title[The Rogue One]{The Rogue One: A Dark Matter-Dominated Dwarf in the Trail of Dark Matter-Deficient Galaxies}
\author[D. Vaz et al.]{Daniel Vaz,$^{1,2}$\thanks{E-mail: ddiasvaz@swin.edu.au}
Arsen Levitskiy,$^{1}$
Duncan Forbes,$^{1}$
Aaron J. Romanowsky,$^{3,4}$
Jonah Gannon,$^{5,1}$
\newauthor Jean P. Brodie,$^{1}$
Yimeng Tang, $^{4}$
Kenji Bekki, $^{6}$
and Warrick J. Couch $^{1}$
\\
$^{1}$ Centre for Astrophysics and Supercomputing, Swinburne University of Technology, Hawthorn, VIC 3122, Australia\\
$^{2}$ Instituto de Astrofísica e Ciências do Espaço, Universidade do Porto, CAUP, Rua das Estrelas, 4150-762 Porto, Portugal\\
$^{3}$ Department of Physics and Astronomy, San José State University, One Washington Square, San Jose, CA 95192, USA \\
$^{4}$ Department of Astronomy and Astrophysics, University of California Santa Cruz, 1156 High Street, Santa Cruz, CA 95064, USA \\
$^{5}$ Department of Astronomy \& Astrophysics, University of Toronto, 50 St. George Street, Toronto, ON M5S 3H4, Canada \\
$^{6}$ International Centre for Radio Astronomy Research (ICRAR), M468, The University of Western Australia, 35 Stirling Hwy, Crawley, WA 6009, Australia
}

\date{Accepted XXX. Received YYY; in original form ZZZ}

\pubyear{\the\year{}}

\begin{document}
\label{firstpage}
\pagerange{\pageref{firstpage}--\pageref{lastpage}}
\maketitle

\begin{abstract}
We present the first measurement of the stellar velocity dispersion of LEDA\,4014647,
a compact dwarf galaxy spatially aligned with the NGC\,1052
trail, a linear chain of dark matter-deficient galaxies (DMDGs) proposed to have formed in a single high-speed dwarf--dwarf collision $\sim 8$\,Gyr ago.
Using integral-field spectroscopy obtained with the Keck Cosmic Web Imager (KCWI), we measure a stellar velocity dispersion of $\sigma = 15.8 \pm 2.0$\,km\,s$^{-1}$ within the
half-light radius ($R_\mathrm{e}$).
We derive a dynamical mass within the 3D half-light radius of $M_\mathrm{dyn}(<r_{1/2}) = (1.5 \pm 0.4) \times 10^8\,M_\odot$, a factor of $\sim 3$ larger than the enclosed stellar mass, implying a DM fraction of $f_\mathrm{DM} = 0.7 \pm 0.1$ at $r_{1/2}$.
Full-spectrum stellar population fitting yields
an old ($t \approx 13$\,Gyr) and metal-poor
($[\mathrm{M/H}] \approx -1.2$\,dex) stellar population, consistent
with independent photometric estimates and predating the proposed
collision epoch. LEDA\,4014647 is therefore a DM-dominated system, and cannot have formed from the DM-free, shock-compressed gas that produced the proposed trail DMDGs DF2, DF4, and DF9.
Its nature, however, remains an open question: while a chance-projection along the trail axis is a possible explanation, its old stellar population, compact morphology, and DM content are also consistent with the surviving collisionless remnant of one of the progenitor dwarfs of the dwarf--dwarf collision.
Distinguishing between these two scenarios will ultimately require deeper spectroscopy to tighten the stellar population and DM halo 
constraints, and targeted  $N$-body and hydrodynamical simulations tracking the post-encounter fate of the progenitors.
\end{abstract}




\section{Introduction}
\label{sec:introduction}

Within the standard $\Lambda$ Cold Dark Matter ($\Lambda$CDM) cosmological framework, dwarf galaxies are predicted to reside within massive dark matter (DM) halos \citep[e.g.][]{Mo_2010, Boylan_Kolchin_2011, Behroozi_2013, Bullock_2017}. This expectation was fundamentally challenged by the discovery of NGC\,1052-DF2 and NGC\,1052-DF4 (hereafter DF2 and DF4), two ultra-diffuse galaxies \citep[UDGs;][see \citealt{Gannon_2026} for a review]{van_Dokkum_2015} associated with the NGC\,1052 group. UDGs are an extreme class of low-surface-brightness dwarf characterised by large effective radii ($R_\mathrm{e} \gtrsim 1.5$\,kpc) and faint central surface brightnesses ($\mu_{0,g} \gtrsim 24$\,mag\,arcsec$^{-2}$), found across a wide range of environments \citep[e.g.][]{Koda_2015, van_der_Burg_2016, Roman_2017, Forbes_2020, La_Marca_2022}. In a departure from the typical DM-dominated nature of UDGs, however, high-resolution kinematic studies of the unusually luminous globular cluster (GC) systems and integrated stellar light of DF2 and DF4 revealed extraordinarily low velocity dispersions ($\sigma \lesssim 10$\,km\,s$^{-1}$), indicating a near-total absence of DM in the inner regions \citep{van_Dokkum_2018, Wasserman_2018, Danieli_2019, van_Dokkum_2019, Emsellem_2019, Shen_2021, Shen_2023}. The origin of these DM-deficient galaxies (DMDGs) has since sparked intense debate. One proposed explanation is extreme tidal stripping by the massive host NGC\,1052, which could, in principle, remove the DM halo while leaving the central stellar body intact \citep{Ogiya_2018, Ogiya_2021, Jackson_2021, Moreno_2022}, though this requires finely tuned orbital configurations, and does not account for the unusually luminous GC systems observed in DF2 and DF4. 
A further suggestion is that DF2 and DF4 may lie at $D \approx 13$\,Mpc \citep{Trujillo_2019, Monelli_2019}, which would substantially revise their inferred masses and potentially eliminate  the DM deficit. Nevertheless, multiple subsequent measurements using independent techniques have consistently favored larger distances. Deep \textit{Hubble Space Telescope} imaging using the tip of the red giant branch (TRGB) method places DF2 at $21.7 \pm 1.2$\,Mpc \citep{Shen_2021, Shen_2023} and DF4 at $20.0 \pm 1.6$\,Mpc \citep{Danieli_2020}. Other approaches yield similarly large values: \citet{Beasley_2025} derived a distance of $D = 16.2 \pm 1.3\,\mathrm{(stat.)} \pm 1.7\,\mathrm{(sys.)}$\,Mpc to DF2, assuming that its GCs follow the same relation between absolute magnitude and velocity dispersion as those of the Milky Way and M31, and a similar recent analysis by van Heumen et al. (submitted) finds $19.1 \pm 4.3$\,Mpc. Finally, recent surface brightness fluctuation (SBF) measurements place the trail galaxies at $D \approx 17$--$20$\,Mpc \citep{Tang_2026}. Taken together, these measurements are consistent with membership of the NGC\,1052 group \citep{Shen_2021} and support the conclusion that these galaxies are indeed DMDGs, even though the uncertainties remain large.

If these galaxies are indeed DM-deficient, an alternative formation pathway is the ``bullet dwarf'' scenario \citep{Silk_2019, Shin_2020}, a dwarf-galaxy-scale analogue of the Bullet Cluster \citep{Clowe_2006}. In a high-speed encounter between two gas-rich, DM-dominated dwarfs, the collisional and collisionless components decouple in phase space: the DM and pre-existing stars continue largely unimpeded along their original trajectories, while the interstellar medium of each progenitor is ram-pressure decelerated, shock-compressed, and left behind. This shock-heated gas then cools and fragments, seeding the formation of massive, coeval GCs \citep{Lee_2021} and, depending on the collision parameters, a variable number of new DMDGs arranged in a linear trail \citep{van_Dokkum_2022, Lee_2024}. This scenario also could reproduce the anomalously large sizes and unusually luminous GC systems of DF2 and DF4 \citep{Silk_2019, Shin_2020, Lee_2021}, addressing several of their unusual properties within a single framework.

The bullet dwarf scenario makes three categories of testable predictions. Firstly, the newly formed DMDGs should geometrically align along the collision axis, with their radial velocities increasing linearly with projected position along the trail, reflecting the gradual deceleration of the dissipating gas filament \citep{van_Dokkum_2022, Lee_2024}. Secondly, in terms of stellar populations, the trail galaxies should exhibit homogeneous, coeval ages consistent with the collision epoch, as their stars formed entirely from the post-shock gas in a single burst; accordingly, they should be DM-deficient by construction, since the DM had already decoupled before star formation commenced. Finally, the model also predicts the existence of surviving progenitor remnants: the two original DM-rich dwarf cores, having experienced minimal hydrodynamical drag, pass through the encounter largely intact and are expected to lie near the leading edges of the diverging trajectory. These remnants are qualitatively distinct from the trail galaxies: they must harbour older, pre-collision stellar populations, retain their DM halos, and may stand out morphologically as higher-surface-brightness objects \citep{van_Dokkum_2022, Lee_2024}.

Observationally, these predictions have been partially confirmed, though tensions remain. \citet{van_Dokkum_2022} demonstrated that the present-day line-of-sight distances and radial velocities of DF2 and DF4 are consistent with joint formation in the aftermath of a single bullet-dwarf collision $\sim 8$\,Gyr ago, and identified a tightly aligned linear substructure of seven to eleven faint galaxies in the NGC\,1052 field.
The structure spans $\sim 380$\,kpc in projection, with an inferred line-of-sight extent of $\sim 2$\,Mpc, and its alignment is statistically unique among wider diffuse galaxy
catalogues \citep{Tanoglidis_2021, Zaritsky_2022, Keim_2025}. \citet{Keim_2025} subsequently provided kinematic support, finding that five of the seven brightest trail galaxies follow a linear velocity--position gradient, with radial velocities increasing systematically with projected distance along the trail. The three-dimensional geometry, however, remains uncertain. Distance estimates for the individual trail members differ between methods and datasets, and their relative line-of-sight ordering is not yet
established \citep{Tang_2026}. Since the collision scenario predicts a specific correspondence between position along the trail and radial velocity, this ordering is required to establish that the trail is a coherent three-dimensional structure rather than a chance alignment in projection.

The stellar populations of the trail members provide further, though not unambiguous, support. Spectroscopic and SED-fitting analyses find that DF2 and DF4 have consistent ages and metallicities \citep{Fensch_2019, Lara_2019, Buzzo_2022, van_Dokkum_2022b}, and the GC systems of both galaxies are nearly monochromatic, spanning a narrow range of colours
\citep{van_Dokkum_2022b}. Extending this comparison to the fainter trail
members, however, reveals that the populations are correlated but not
identical, with a possible trend towards younger ages at lower masses \citep{Tang_2025}. Nevertheless, the age uncertainties are large and the measurements remain mutually consistent.
In addition, the GCs of DF2 have been found to span a wide range in
metallicity and to be predominantly more metal-poor than the diffuse
stellar body \citep{Fensch_2019, Fahrion_2025}. This is not fully
consistent with a population formed in a single event from the available
gas, which would be expected to share a common metallicity, but more data is needed.

Independent support for this underlying mechanism comes from outside the
NGC\,1052 group. In the Fornax cluster, FCC\,224 was first identified as a candidate DMDG based on its apparently overluminous GC population \citep{Romanowsky_2024, Tang_2025_224}. Subsequent follow-up has shown that FCC\,224 is genuinely DM-deficient and has further revealed a nearby companion candidate DMDG (FCC\,240), forming a possible analogue to the paired DMDG phenomenology seen in DF2 and DF4 \citep{Buzzo_2025, Buzzo_2026}, and importantly, with robust distance estimates that strengthen the case for their DM deficiency. Taken together, these emerging systems suggest that collision-driven pathways to DMDG formation are viable and may not be confined to the NGC\,1052 group.
The NGC\,1052 trail therefore remains a plausible but unconfirmed physical structure.

If this formation pathway is correct, a key prediction is that the other trail galaxies in the NGC\,1052 group, having formed from the decoupled gas, should also be severely deficient in DM. Measuring the dynamical masses of these objects is therefore crucial to test this scenario. This is, however, an observationally demanding task given the exceptionally faint and diffuse nature of these targets. Recently, \citet{Keim_2026} successfully obtained stellar kinematics for NGC\,1052-DF9 (hereafter DF9), confirming it as the third DM-deficient galaxy within the trail, a discovery that strongly bolsters the hypothesis that the remaining trail members share a similar DM deficiency.

In this paper, we continue to test the bullet dwarf scenario, focusing on another galaxy spatially aligned with the proposed trail: LEDA 4014647 (hereafter LEDA). After DF2, DF4, and DF9, LEDA is the brightest galaxy possibly associated with the structure, and therefore the most observationally accessible remaining target for dynamical mass measurements. It shares a variety of structural and photometric properties with DF9, including nearly identical stellar mass, S\'ersic index, and axis ratio, with both galaxies also hosting a nuclear star cluster \citep{Tang_2025}. Despite these similarities, LEDA stands out as a distinct physical extreme among the galaxies associated with the trail: it is significantly older (from SED fitting) and more compact than any other galaxy in the structure \citep{Tang_2025}.

Given its relative brightness and compact morphology, \citet{van_Dokkum_2022} initially identified LEDA as a candidate interloper, suggesting it might be a galaxy unrelated to the trail. This exclusionary classification was seemingly corroborated by \citet{Keim_2025}, who measured a systemic velocity of $cz = 1596^{+3}_{-2}$\,km\,s$^{-1}$ for LEDA and flagged it as an interloper on the grounds that this velocity deviates by $>100$\,km\,s$^{-1}$ from the linear kinematic trend predicted at LEDA's projected position along the trail. However, simulations of the bullet dwarf collision \citep{Lee_2024} explicitly showed that a substantial line-of-sight velocity scatter is naturally expected for galaxies within the trail, meaning such an offset does not preclude membership.

The nature of LEDA therefore remains an open question: is it a chance-projection interloper unrelated to the DMDGs, a genuine trail DMDG, or the surviving remnant of one of the two progenitor dwarfs that produced the entire structure? The key to distinguishing between these scenarios lies in its DM content. A galaxy formed from the shock-compressed gas, like DF2, DF4, and DF9, should be DM-deficient; a progenitor remnant or chance interloper should retain a DM halo. In this paper, we present the first measurement of the stellar velocity dispersion of LEDA from integral field spectroscopy, from which we derive its dynamical mass and DM content, and discuss the implications for its role in the NGC\,1052 bullet dwarf collision.

This paper is organised as follows. In Section~\ref{sec:data}, we describe the observations and data reduction. We present our kinematic and stellar population measurements in Section~\ref{sec:results}. In Section~\ref{sec:discussion}, we discuss the implications for LEDA's dynamical state and its nature in the context of the collision scenario. We summarise our conclusions in Section~\ref{sec:conclusion}.


\section{Data} \label{sec:data}

LEDA is a compact dwarf in the NGC\,1052 group whose photometric and structural properties have been characterised by \citet{Tang_2025} through SED fitting, yielding stellar mass, age, metallicity, and morphological parameters.
A systemic velocity of $cz = 1596^{+3}_{-2}$\,km\,s$^{-1}$ was measured 
by \citet{Keim_2025} from Keck/LRIS spectroscopy. The properties adopted throughout this work are summarised in Table~\ref{tab:leda_properties}.

\begin{figure*}
\centering
\includegraphics[width=\textwidth]{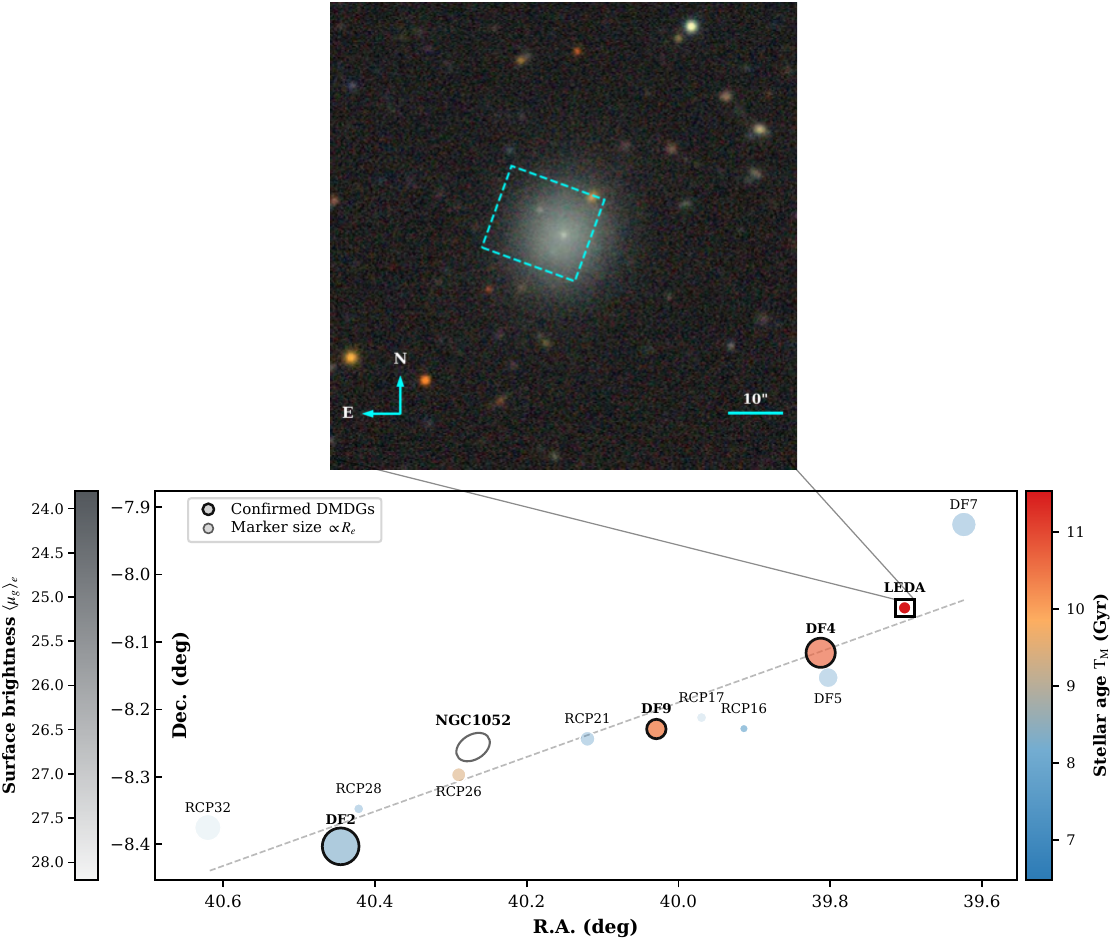}
    \caption{Map of the NGC\,1052 galactic trail. The marker radius for each galaxy is scaled proportional to its physical effective radius ($R_\mathrm{e}$), coloured according to its mass-weighted age, and its opacity scales with surface brightness ($\langle\mu_g\rangle_e$), using values compiled from \citet{Tang_2025} and \citet{Keim_2025}. The position and orientation of the massive elliptical galaxy NGC\,1052 are indicated by an open ellipse, whose shape and position angle reflect the galaxy's approximate morphology. LEDA is visually distinct as the oldest, most compact, and highest surface brightness object in the trail, situated at the western extremity near DF7. The dashed line shows a linear fit to the trail members to guide the eye. The zoom-out panel shows a $grz$ colour composite image of LEDA from the Legacy Survey DR10 \citep{Dey_2019}. The dashed rectangle outlines the $16.5\arcsec \times 20.4\arcsec$ field of view of the KCWI (Medium slicer). The $10\arcsec$ scale bar corresponds to a physical distance of 860\,pc at the adopted distance. North is up and East is to the left.}
\label{fig:fov}
\end{figure*}

For visual reference, the spatial distribution of the putative trail members across the NGC\,1052 field is shown in Fig.~\ref{fig:fov}. In this map, the symbols are scaled in area by their physical effective radius ($R_\mathrm{e}$), colour-coded by their mass-weighted age, and their opacity is scaled by their surface brightness. As is immediately apparent, LEDA stands out from the rest of the trail as the oldest, most compact, and highest surface brightness member of the structure. Spatially, it is situated towards the extreme western edge of the trail, second only to the UDG NGC\,1052-DF7 (hereafter DF7). The zoom-out panel shows a $grz$ colour composite image of LEDA from the Legacy Survey DR10 \citep{Dey_2019}, with the KCWI mosaic field of view overlaid.

\subsection{Observations}
\label{sec:observations}

The integral field spectroscopy data for LEDA were obtained using the Keck Cosmic Web Imager \citep[KCWI;][]{Morrissey_2018} on the 10\,m Keck\,II telescope at Maunakea, Hawai'i. The observations were carried out on the night of 2024 November 6 (Programme ID: W285, PI: Gannon). We used the Medium slicer with the high-resolution Blue grating (BH3), providing a spectral coverage from 4830\,\AA\ to 5330\,\AA. This wavelength range covers the key stellar absorption features required for kinematic and stellar population analysis, including the Balmer H$\beta$ line, the \ion{Fe}{i}\,$\lambda$5015 feature, and the Mg\,$b$ triplet. We acquired four individual science exposures of 1800\,s each, totalling 2\,hr of on-target integration. The average seeing during the observations was approximately $1.5\arcsec$, under clear sky conditions.

\begin{table}
\centering
\caption{Properties of LEDA\,4014647 adopted from the literature.}
\label{tab:leda_properties}
\renewcommand{\arraystretch}{1.4}
\begin{tabular}{lc}
\hline
\hline
Property & Value \\
\hline
RA (J2000)$^a$ & 39.70206 \\
Dec. (J2000)$^a$ & $-$8.04938 \\
Distance ($D$)$^b$ & $17.7 \pm 1.4$ Mpc \\
Effective Radius ($R_\mathrm{e}$)$^a$ & $7.53 \pm 0.05\arcsec$ ($\sim 646$ pc) \\
Axis Ratio ($b/a$)$^a$ & $0.866 \pm 0.006$ \\
Mean Surface Brightness ($\langle \mu_g \rangle_e$)$^a$ & $23.80 \pm 0.02$ mag arcsec$^{-2}$ \\
Stellar Mass ($\log(M_*/M_\odot)$)$^a$ & $7.98 \pm 0.06$ \\
Metallicity ([M/H])$^a$ & $-1.38^{+0.11}_{-0.14}$ \\
Mass-weighted Age$^a$ & $11.54^{+1.12}_{-1.69}$ Gyr \\
Systemic Velocity ($V$)$^c$ & $1596^{+3}_{-2}$\,km\,s$^{-1}$ \\
\hline
\multicolumn{2}{l}{$^a$\citet{Tang_2025}; $^b$\citet{Tang_2026}; $^c$\citet{Keim_2025}.} \\
\end{tabular}
\end{table}

\subsection{Data Reduction}
\label{sec:reduction}

The raw data were reduced using the standard KCWI Data Reduction Pipeline \citep[KCWI\_DRP;][]{Morrissey_2018}, which performs bias subtraction, flat-fielding, cosmic ray removal, wavelength calibration, and illumination correction, with the exception of the pipeline's built-in background subtraction, which was omitted in favour of the custom sky subtraction procedure described below. We further applied cube trimming and residual gradient removal to each individual reduced cube following the procedures described in \citet{Gannon_2020}. The trimmed cubes were then reprojected and mosaic stacked using \textsc{MontagePy}\footnote{\url{https://github.com/Caltech-IPAC/Montage/tree/main/python/MontagePy}} \citep{Berriman_2017} to produce the final full datacube.

The instrumental spectral resolution was characterised from isolated atmospheric airglow emission lines in the science frames, specifically the [\ion{N}{i}]\,$\lambda\lambda$5198, 5200 doublet, and cross-checked against arc lamp calibration exposures. This yields a line-spread function well described by a Gaussian with $\sigma_{\rm inst} \approx 12.2$\,km\,s$^{-1}$.

Given the low surface brightness of LEDA, accurate sky subtraction is essential. We identified a blank, source-free region within the science field of view and used it to construct a local sky model. We first fitted the prominent sky emission lines to account for spatial flux variations across the field, then generated a median sky spectrum from the blank region and subtracted it spaxel-by-spaxel from the full datacube.

For the primary kinematic and stellar population analysis, we collapsed the sky-subtracted datacube within a circular aperture of $1\,R_\mathrm{e}$ ($= 7.53\arcsec$), yielding an integrated spectrum with an empirical signal-to-noise ratio (SNR) of $\approx 15$ per pixel ($0.25$\,\AA\,pixel$^{-1}$). To search for evidence of ordered rotation, we additionally extracted spectra from pairs of semi-circular apertures obtained by bisecting the $1\,R_\mathrm{e}$ aperture along multiple position angles, following \citet{Levitskiy_2025}, and compared the mean line-of-sight velocities of the two opposing halves at each angle.

\section{Results}\label{sec:results}

\subsection{Stellar Kinematics}
\label{sec:kinematics}

\begin{figure*}
    \centering
    \includegraphics[width=\textwidth]{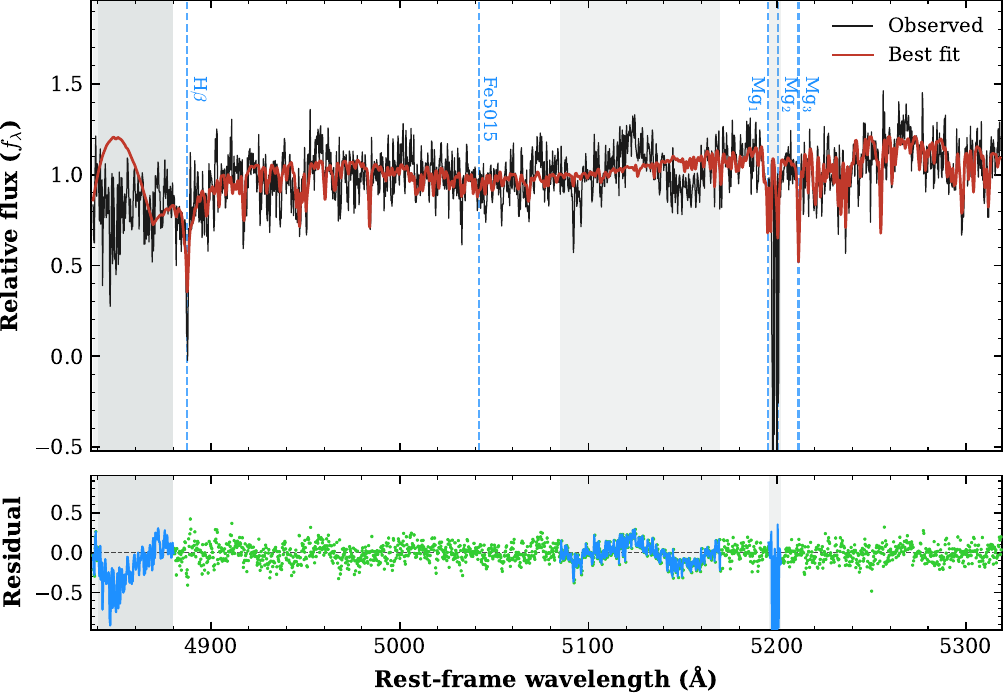}
    \caption{%
        \textsc{pPXF} fit to the KCWI blue-side spectrum of LEDA
        extracted within the half-light radius $R_\mathrm{e}$.
        \textit{Upper panel:} Observed spectrum (black) and best-fit
        \textsc{pPXF} model (red). Dashed blue vertical lines mark
        key stellar absorption features: H$\beta$\,$\lambda$4861,
        \ion{Fe}{i}\,$\lambda$5015, and the Mg\,$b$ triplet
        ($\lambda$5167, 5173, 5184\,\AA).
        \textit{Lower panel:} Fit residuals (observed minus model).
        Green points mark the fitted pixels; blue segments indicate
        spectral regions excluded from the fit.
    }
    \label{fig:specfit_Re}
\end{figure*}

\subsubsection{Analysis}

The stellar kinematics of LEDA were derived from the
extracted integrated spectra using the Penalised Pixel-Fitting code
\citep[\textsc{pPXF};][]{Cappellari_2004, Cappellari_2017, Cappellari_2023}.
\textsc{pPXF} fits a linear combination of stellar template spectra
convolved with a parametric line-of-sight velocity distribution,
modelled here as a Gaussian characterised by the systemic velocity $V$ and
velocity dispersion $\sigma$.
Additive and multiplicative Legendre polynomials of degree 8 were
included in the fit to account for continuum shape differences between the
galaxy spectrum and the templates, likely arising from residual flux
calibration uncertainties and template mismatch. We tested a range of
configurations, including using only additive, and varying the degree from 4 to 12; 
while sub-optimal choices introduce localized, systematic residuals specifically around the cores and wings of the main absorption lines (the final residuals of our adopted fit are shown in Figure~\ref{fig:specfit_Re}), the recovered kinematic parameters remain robust across all tested configurations.

We used the high-resolution theoretical stellar spectra
of \citet{Coelho_2014}, with a spectral resolution of $\mathrm{FWHM} = 0.02$\,\AA, well below the
instrumental resolution of the observations (see Section~\ref{sec:reduction}).
The fit encompasses the full available wavelength range, with the exception
of spectral regions masked due to residual sky subtraction artefacts,
covering key stellar absorption features including the Balmer H$\beta$ line,
the \ion{Fe}{i}\,$\lambda$5015 feature, and the Mg\,$b$ triplet.
We note that the choice of spectral mask does not affect the kinematic results. We tested fits with various masking strategies, including no mask, masking only the H$\beta$ region, and the full mask adopted here, and found that the recovered kinematics remained consistent in all cases.
The best-fit \textsc{pPXF} solution is shown in Fig.\,\ref{fig:specfit_Re}.

Uncertainties on the kinematic parameters were estimated using a bootstrap approach \citep{Cappellari_2023}, in which $N_\mathrm{boot} = 100{,}000$ realisations of the spectrum were generated by multiplying the best-fit residuals by independent random sign-flip vectors.
Each realisation was re-fitted with \textsc{pPXF}, and the quoted
uncertainties correspond to the 16th--84th percentile range
of the resulting bootstrap distributions, which account for correlated
noise and non-Gaussian residuals independently of the formal covariance
matrix of the fit.
The full bootstrap distributions are shown in
Appendix~\ref{app:bootstrap_corner}.

\subsubsection{Results}
\label{sec:kin_results}

From the spectrum extracted within the effective radius $R_\mathrm{e}$,
we measure a barycentric systemic velocity of $V = 1600.7 \pm 1.3$\,km\,s$^{-1}$, consistent with the value reported by \citet{Keim_2025}, and a stellar velocity dispersion of $\sigma = 15.8 \pm 2.0$\,km\,s$^{-1}$.
The goodness-of-fit metric $\chi^2/\nu = 2.39$, where $\nu$ denotes
the number of degrees of freedom, 
indicates an acceptable solution, with the modest excess above unity
likely reflecting the low SNR
of the data or low-level residual systematics
in the continuum rather than a failure of the kinematic model.
We note that we do not correct the measured velocity dispersion for stellar atmospheric broadening (internal rotation and macroturbulence) or binary orbital motions. Both contributions are expected to be small\footnote{If we conservatively adopt upper limits of $\sigma_{\rm broad} = 5.0\,{\rm km\,s^{-1}}$ for atmospheric broadening and $\sigma_{\rm bin} = 1.0\,{\rm km\,s^{-1}}$ for binaries, this yields a maximum correction of $\sim 5\%$. This is well within our measurement uncertainties, and since the true value is unknown, no correction is applied.} relative to our measured $\sigma$, with the atmospheric term subdominant \citep{Shen_2023} and binary effects negligible in old stellar systems \citep[e.g.][]{Vaz_2026}.

To search for ordered rotation, we additionally fitted spectra
extracted from a series of semi-apertures at different position angles,
following the approach described in Section~\ref{sec:data}.
The recovered line-of-sight velocities were consistent with one another
within the bootstrap uncertainties across all position angles,
providing no significant evidence for rotation in LEDA.
However, given the high axis ratio ($b/a = 0.866$; \citealt{Tang_2025}), 
LEDA may be viewed nearly face-on, meaning intrinsic rotation could simply be hidden by projection effects.
We discuss the implications of this result, together with the measured
velocity dispersion, for the dynamical state and formation history of
LEDA in Section~\ref{sec:discussion}.


\subsection{Stellar Populations}
\label{sec:populations}

\subsubsection{Analysis}

The stellar population properties of LEDA were derived from the same
$R_\mathrm{e}$-aperture spectrum used for the kinematic analysis
(Section~\ref{sec:kinematics}), using \textsc{pPXF} in its regularised
full-spectrum fitting mode.
In this configuration, \textsc{pPXF} fits a non-negative linear combination
of simple stellar population (SSP) templates over a two-dimensional grid
of age and metallicity, with a regularisation parameter $\lambda$ controlling
the smoothness of the recovered star formation history \citep{Cappellari_2023}.
The systemic velocity and velocity dispersion were fixed to the values
derived in Section~\ref{sec:kinematics}, leaving the stellar population
weights as the only free parameters alongside the multiplicative and
additive polynomials of degree 8. As in Section~\ref{sec:kinematics}, we tested a 
range of configurations, including using only additive or only 
multiplicative polynomials and varying the degree from 4 to 12; the 
recovered stellar population parameters remain robust across all tested 
configurations, with the residuals exhibiting the same localized behavior discussed in Section~\ref{sec:kinematics}.

As a template library we used the E-MILES SSP models \citep{Vazdekis_2015},
computed with BaSTI isochrones \citep{Hidalgo_2018}, a \citet{Kroupa_2001}
universal initial mass function, and scaled-solar abundance ratios.
The models have a spectral resolution of \mbox{$\mathrm{FWHM} = 2.52$\,\AA} and
span ages between \mbox{0.03} and \mbox{14\,Gyr} and metallicities between
\mbox{$[\mathrm{M/H}] = -2.42$} and \mbox{$[\mathrm{M/H}] = +0.40$\,dex}.

From the recovered weight distribution, we derive the mass-weighted mean age ($t_M$) and metallicity ($[\mathrm{M/H}]$) following \citet{Cappellari_2023}, converting the luminosity-weighted \textsc{pPXF} output into mass fractions by dividing each template weight by its mass-to-light ratio. The cumulative star formation history (SFH) is then constructed from the ordered mass fractions as a function of lookback time.
The regularisation parameter was set to $\lambda = 10$, chosen such
that the increase in $\chi^2$ relative to the unregularised solution
is consistent with the expected statistical fluctuations given the
number of fitted pixels \citep{Cappellari_2023}.
We tested various values of $\lambda$ and found that while higher regularisation predictably produces a smoother and more extended star formation history, the bootstrap procedure described below drives all solutions towards the same preferred values, rendering the results insensitive to the specific choice of $\lambda$.

Uncertainties were estimated using the same wild bootstrap procedure
described in Section~\ref{sec:kinematics}, with $N_\mathrm{boot} = 100{,}000$
realisations. For $t_M$ and $[\mathrm{M/H}]$, the reported values correspond to the mode of the bootstrap
distributions, estimated via kernel density estimation, with asymmetric
uncertainties defined by the 68.3 per cent Highest Density Interval (HDI), the shortest interval containing 68.3 per cent of the bootstrap realisations. The cumulative SFH shown in Fig.~\ref{fig:sfh} is the median of the $N_\mathrm{boot}$ realisations, with the shaded uncertainty band corresponding to the HDI centred on the mode. The lookback times $t_{50}$ and $t_{90}$, at which 50 and 90 per cent of the stellar mass had been assembled, are read directly from the median SFH curve. As an additional robustness check, we also extract the modes of the $t_{10}$, $t_{50}$, and $t_{90}$ distributions from the individual bootstrap realisations; the resulting violin plots are shown in Fig.~\ref{fig:violin_plots} (Appendix~\ref{app:bootstrap_corner}). The full bootstrap distributions are shown in
Appendix~\ref{app:bootstrap_corner}.

\subsubsection{Results}
\label{sec:pop_results}

\begin{figure}
    \centering
    \includegraphics[width=\columnwidth]{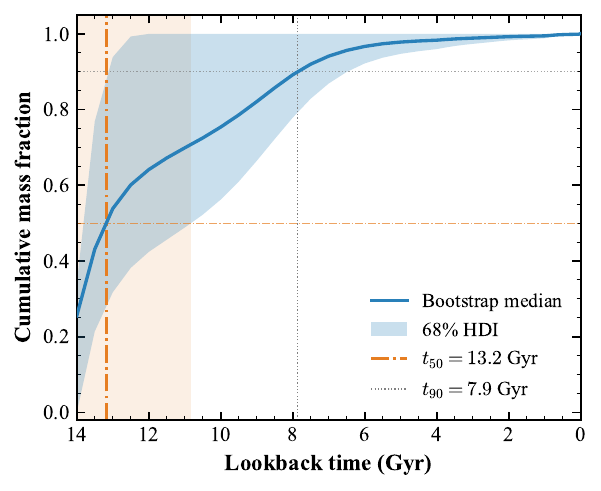}
    \caption{%
        Cumulative mass-weighted star formation history of LEDA,
        recovered from the \textsc{pPXF} fit to the $R_\mathrm{e}$-aperture
        KCWI blue-side spectrum using E-MILES SSP templates.
        The solid blue line shows the median cumulative star formation
        history from $N_\mathrm{boot} = 100{,}000$ wild bootstrap
        realisations; the shaded blue region indicates the
        68 per cent HDI centred on the mode.
        The dash-dotted orange vertical line marks the lookback time
        at which 50 per cent of the stellar mass had formed ($t_{50}$),
        with the shaded orange band indicating its bootstrap
        uncertainty.
        The dotted grey line marks the lookback time at which
        90 per cent of the stellar mass had formed ($t_{90}$).
        The $x$-axis is oriented so that lookback time decreases
        to the right, placing older stellar populations on the left.
    }
    \label{fig:sfh}
\end{figure}

The recovered mass-weighted mean age is $t_M = 13.2^{+0.4}_{-3.0}$\,Gyr, and the mean metallicity is
$[\mathrm{M/H}] = -1.22 \pm 0.32$\,dex.
The half-mass formation time read from the median cumulative SFH curve (Fig.~\ref{fig:sfh}) is $t_{50} = 13.2$\,Gyr, in total agreement with $t_M$; we therefore adopt $t \equiv t_M \approx t_{50} = 13.2^{+0.4}_{-3.0}$\,Gyr as a single representative age throughout.
The cumulative SFH curve provides a smoothed representation of the mass assembly, from which one might infer a continuous formation history extending up to $t_{90} \approx 8$\,Gyr. However, to better characterise the robustness of these results, we analysed the star formation history of each bootstrap iteration individually (see Appendix~\ref{app:bootstrap_corner}). Extracting the $t_{10}$, $t_{50}$, and $t_{90}$ values directly from the modes of the individual bootstrap realisations (Fig.~\ref{fig:violin_plots}) reveals a strong underlying preference for a single, extremely old burst of star formation. Nevertheless, some iterations do exhibit a more extended SFH, as indicated by the $t_{90}$ distribution: approximately half of the iterations return 
$t_{90} \geq 8$\,Gyr, with the remainder extending to later times 
($\sim 25\%$ with $t_{90} < 6$\,Gyr). This variance, confined to the tail of the SFH, highlights the expected degeneracy inherent in fitting a low-SNR spectrum. Despite this degeneracy, the results consistently demonstrate that LEDA is a particularly ancient system, with the bulk of its star formation having ceased at an epoch consistent with the timing of 
the putative NGC\,1052 collision.

We caution, however, that the uncertainties on these measurements are
large, as reflected in the asymmetric bootstrap confidence intervals.
The blue-side KCWI spectrum of LEDA has a
SNR of only $\approx 15$ per pixel, and within the observed wavelength
range the spectrum contains limited stellar population discriminating
power, being dominated in practice by the H$\beta$ absorption line.
Furthermore, while the choice of spectral mask slightly affects the initial unperturbed fit, the bootstrap procedure robustly converges to a preferred solution that is consistently old, as clearly demonstrated by the asymmetric uncertainties strongly tailing towards younger ages but sharply peaking at the oldest templates.
As an additional robustness check, we performed the stellar population fits to a spectrum extracted from the central nucleus of LEDA and to a spectrum extracted from the extended stellar body within our $1\,R_\mathrm{e}$ footprint (masking the nucleus). The recovered parameters in both cases were fully compatible within $1\sigma$, indicating no significant stellar population gradients or central anomalies.

The stellar population parameters derived here are broadly consistent with those reported by \citet{Tang_2025}, who found LEDA to be old 
($t_M = 11.54^{+1.12}_{-1.69}$\,Gyr) and metal-poor 
($[\mathrm{M/H}] = -1.38^{+0.11}_{-0.14}$\,dex) based on SED fitting. Both analyses converge on a picture of an old, metal-poor stellar system that stands out clearly from the other
trail members.

Furthermore, with $[\mathrm{M/H}] \approx -1.22$\,dex, LEDA is
among the most metal-poor galaxies along the trail.
Although this central value falls slightly below the predictions of the 
mass--metallicity relation derived by \citet{Tang_2025} for the trail dwarfs  ($\approx -1.19$\,dex) and the Local Group relation from \citet{Simon_2019} ($\approx -1.1$\,dex), it remains formally consistent with both once the  observational uncertainties are taken into account (see \citealt{Tang_2025} for a broader overview of the trail galaxies). 
Such low metallicities are characteristic of systems that formed
early and rapidly, before significant metal enrichment, and are
more typical of ancient dwarf spheroidals.
Taken together, the stellar population properties of LEDA are
inconsistent with membership of the trail as a collision product.


\begin{table*}
\centering
\caption{Measured and derived properties of LEDA\,4014647 from this work.
From left to right, the columns show the systemic velocity $V$,
stellar velocity dispersion within $R_\mathrm{e}$ ($\sigma$),
dynamical mass within the 3D deprojected half-light radius
$M_\mathrm{dyn}(<r_{1/2})$ derived
using the \citet{Wolf_2010} estimator,
dynamical-to-stellar mass ratio $M_\mathrm{dyn}/M_\star$,
DM fraction $f_\mathrm{DM}$,
dynamical mass-to-light ratio $(M/L)_\mathrm{dyn}$,
and the mass-weighted mean age $t$ and metallicity $[\mathrm{M/H}]$
from full-spectrum \textsc{pPXF} fitting.}
\label{tab:leda_derived}
\renewcommand{\arraystretch}{1.2}
\begin{tabular}{cc|cccc|cc}
\hline\hline
$V$ & $\sigma$ &
$M_\mathrm{dyn}(<r_{1/2})$ & $M_\mathrm{dyn}/M_\star(<\!r_{1/2})$ & $f_\mathrm{DM}$ & $(M/L)_\mathrm{dyn}$ & $t$ & $[\mathrm{M/H}]$ \\
(km\,s$^{-1}$) & (km\,s$^{-1}$) &
($10^8\,M_\odot$) & & & ($M_\odot\,L_\odot^{-1}$) &
(Gyr) & (dex) \\
\hline
$1600.7 \pm 1.3$ & $15.8\pm2.0$ &
$1.5\pm0.4$ & $3\pm 1$ & $0.7\pm0.1$ & $7\pm 3$ &
$13.2^{+0.4}_{-3.0}$ & $-1.22\pm0.32$ \\
\hline
\end{tabular}
\end{table*}

\section{Discussion}\label{sec:discussion}

In this section we use the kinematic and stellar population measurements
of Section~\ref{sec:results} to derive the dynamical mass and DM
content of LEDA, and discuss the implications for its nature in the context
of the NGC\,1052 bullet dwarf collision. A consolidated summary of all
measured and derived quantities is provided in Table~\ref{tab:leda_derived}
for reference.

\subsection{Stellar Kinematics and Dynamical Mass}
\label{sec:disc_kinematics}

We present the first direct measurement of the stellar velocity
dispersion of LEDA, $\sigma = 15.8 \pm 2.0$\,km\,s$^{-1}$.
We find no evidence for ordered rotation from our analysis of
spectra extracted across a series of position angles
(Section~\ref{sec:kin_results}), indicating that the stellar
kinematics are consistent with a purely dispersion-supported
system ($V_\mathrm{rot}/\sigma \approx 0$).

Therefore, using the mass estimator of \citet{Wolf_2010}, which is
particularly well suited to dispersion-supported systems and
minimises sensitivity to orbital anisotropy,
\begin{equation}
    M_{1/2} = \frac{4\,\sigma^2\,R_\mathrm{e}}{G}
    \approx 930\,\left(\frac{\sigma}{\mathrm{km\,s^{-1}}}\right)^2
    \left(\frac{R_\mathrm{e}}{\mathrm{pc}}\right)\,M_\odot,
\end{equation}
\noindent we derive a total mass within the 3D deprojected half-light
radius ($r_{1/2}$) of
\begin{equation}
    M_\mathrm{dyn}(<r_{1/2}) = (1.5 \pm 0.4) \times 10^8\,M_\odot,
\end{equation}
\noindent adopting $\sigma = 15.8$\,km\,s$^{-1}$ and
a effective radius $R_{\mathrm{e,circ}} = R_\mathrm{e}\sqrt{b/a} = 601 \pm 48$\,pc
(corresponding to $7.0 \pm 0.1\arcsec$ at $D = 17.7 \pm 1.4$\,Mpc;
$b/a = 0.866$).

To assess the DM content, we compare the dynamical mass to the
stellar mass within $r_{1/2}$, adopting $M_\star(<r_{1/2}) = M_\star/2$ (see Table \ref{tab:leda_properties}).
The resulting dynamical-to-stellar mass ratio within $r_{1/2}$ is
$M_\mathrm{dyn}/M_\star = 3 \pm 1.0$, implying a
DM fraction of $f_\mathrm{DM} = 0.7 \pm 0.1$ within
the 3D half-light radius, and a dynamical mass-to-light ratio of
\begin{equation}
    (M/L)_{\mathrm{dyn}}(<r_{1/2}) = 7 \pm 3\,M_\odot\,L_\odot^{-1}.
\end{equation}
Although these calculations depend on the assumed distance, the 
associated uncertainties are fully propagated, and the DM-dominated 
nature of LEDA remains robust across any plausible distance 
variation. This confirms that LEDA is a DM-dominated system, and 
does not belong to the population of DMDGs produced in the 
bullet dwarf collision.

\begin{figure*}
    \centering
    \includegraphics[width=\textwidth]{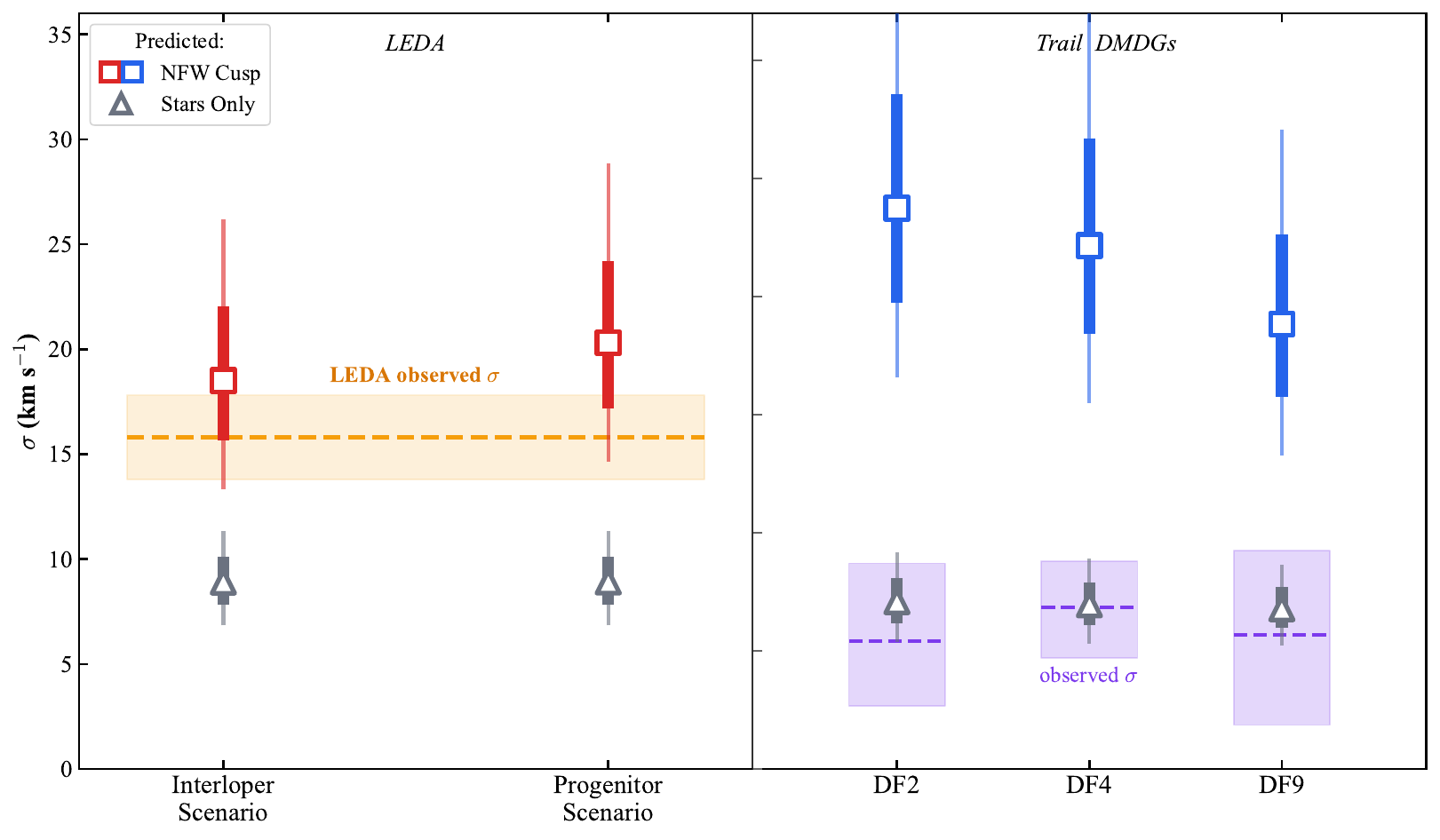}
    \caption{Comparison of the observed velocity dispersions of LEDA 
and the three trail DMDGs (DF2, DF4, and DF9) against dynamical 
models assuming either a stars-only (DM-free) mass budget or a 
full DM halo following a NFW profile \citep{NFW_1997}, with halo 
masses drawn from the SMHM relation of \citet{Moster_2013} via Monte Carlo sampling. The left panel shows the predictions for LEDA under both the interloper ($z \sim 0$) and 
progenitor ($z \sim 1$) scenarios, with the measured velocity 
dispersion indicated as an orange dashed line with yellow shading 
denoting the $1\sigma$ uncertainty. The right panel shows the 
equivalent analysis for the three trail DMDGs, serving as a 
comparison baseline, with their observed velocity dispersions 
similarly shown as dashed lines with shaded $1\sigma$ bands. 
In both panels, stars-only (triangles) and NFW halo (squares) 
predictions are shown as points, with thick and thin bars 
encompassing the 68\% ($1\sigma$) and 95\% ($2\sigma$) Monte Carlo 
confidence intervals, respectively, and the marker indicating the 
median. The two panels reveal a striking contrast: whereas the 
stars-only model reproduces the kinematics of DF2, DF4, and DF9, confirming their DM-deficient nature, it fails entirely for 
LEDA, which requires a DM halo to account for its 
observed velocity dispersion, unambiguously establishing it as a 
DM-dominated system. LEDA's measured velocity dispersion is 
consistent with both the $z \sim 0$ and $z \sim 1$ NFW predictions 
at the $1\sigma$ level, reflecting the large degeneracies in NFW 
concentration at higher redshifts \citep{Dutton_2014}.}

    \label{fig:forest_plot}
\end{figure*}

\subsection{Implications for LEDA's Nature}
\label{sec:disc_implications}

LEDA's DM-dominated dynamical mass and old stellar population, 
predating the proposed collision epoch, immediately exclude it from the 
population of DMDGs. The nature of LEDA and its relationship to the putative trail are therefore called into question, and two 
alternative hypotheses emerge: (a) LEDA is a chance-projection interloper, 
unrelated to the DMDGs regardless of whether the trail is a genuine 
physical structure, or (b) LEDA is a surviving remnant of one of the 
two colliding progenitor dwarfs.

The interloper scenario is, on its face, the most parsimonious: an unrelated galaxy carries no physical requirements beyond coincidental alignment in projection with the DMDGs.
However, this result neither supports nor disproves the bullet dwarf scenario. Within that 
scenario, \citet{van_Dokkum_2022} and \citet{Keim_2025} estimate that $2 \pm 2$ interlopers are expected along the trail by chance-projection alone, and LEDA's brighter, more compact morphology together with its kinematic offset from the trail's velocity gradient were among the earliest indications that it might not belong to the DMDG population. However, as shown by \citet{Lee_2024}, the line-of-sight velocity scatter expected within the trail is substantial, particularly for progenitor remnants whose collisionless components can be gravitationally captured by NGC\,1052 itself; the velocity offset alone therefore does not rule out trail membership. The progenitor hypothesis thus remains open, and would, if supported,  constitute independent evidence in favour of the bullet dwarf 
collision.
Can it be excluded on other grounds? We examine this question below.

\subsection{Unveiling the Progenitor}
\label{sec:disc_progenitor}

If LEDA is indeed a surviving progenitor remnant, its present-day properties should be consistent with the expectations of the bullet-dwarf collision scenario \citep{Silk_2019, Shin_2020, van_Dokkum_2022, Lee_2024}. We evaluate this consistency across four independent lines of evidence: its star formation history, its baryonic budget at the collision epoch, its DM halo properties, and its spatial position within the trail.
For reference, the stellar masses, ages, and $R_\mathrm{e}$ of the trail galaxies used throughout this section are compiled from \citet{Tang_2025}, while their distances are from \citet{Tang_2026}.

\subsubsection{Quenching and Star Formation History}
\label{sec:disc_sfh}

The estimated collision epoch is $z \sim 1$, approximately 8\,Gyr ago \citep{van_Dokkum_2022}. In this scenario, two gas-rich dwarfs collided and were stripped of their entire gas reservoirs, while their stars and DM passed through largely unaffected. A surviving progenitor remnant should therefore show clear evidence of quenching at this epoch: the bulk of its stellar mass should have formed before the collision, with star formation ceasing as the gas supply was removed.

LEDA's measured star formation history is consistent with this picture. Its mass-weighted age of $t \approx 13.2$\,Gyr indicates that the bulk of its stellar mass was already in place well before the proposed collision. The recovered star formation history favours a single early, rapid burst of formation, but the data also permit an extended tail of star formation with a median quenching time of $t_{90} \approx 8$ Gyr ago, which is consistent with a system quenched by the collision itself. We note, however, that the large spread in the $t_{90}$ distribution renders the exact quenching epoch highly uncertain.

\subsubsection{Baryonic Budget}
\label{sec:disc_gas}

Under the bullet-dwarf collision scenario, the stellar mass now distributed across the trail galaxies was formed from gas stripped from the two colliding progenitors. We can therefore use the observed stellar masses of the trail members to estimate the gas content of the progenitors at the time of the collision.

Summing the present-day stellar masses of the trail members with photometric mass estimates from \citet{Tang_2025}, and excluding LEDA as a candidate progenitor remnant, yields a total of $M_{\star,\rm trail}~\approx 5 \times 10^8\,{\rm M_\odot}$. Since the trail stars formed at approximately the collision epoch ($z \sim 1$, $\sim 8$\,Gyr ago), this present-day value must be corrected for the cumulative stellar mass loss that has occurred since formation. We model this correction using the surviving stellar mass fraction ($f_{\rm surv}$) analytical parameterization from the Differentiable Stellar Population Synthesis framework \citep[DSPS;][]{Hearin_2023}, which is explicitly calibrated to emulate FSPS stellar population models \citep{Conroy_2010} with a \citet{Kroupa_2001} initial mass function (consistent with the priors in \citealt{Tang_2025}). Assuming a stellar age of 8\,Gyr (from $z \sim 1$ to today), $f_{\rm surv}(8\,{\rm Gyr}) \approx 0.60$, giving the initial stellar mass formed in the trail as $M_{\star,\rm trail}(z \sim 1) \approx 5 \times 10^8 / 0.60 \approx 8.3 \times 10^8\,{\rm M_\odot}$. We note that if DF7 is also excluded from the sum as a candidate interloper (see Section~\ref{sec:disc_spatial}), the total trail mass decreases only marginally, given DF7's low stellar mass ($\log M_\star/M_\odot = 7.02$; \citealt{Tang_2025}).

However, this stellar mass does not correspond to the entirety of the gas budget, as not all of the stripped gas was converted into stars. Simulations of the bullet-dwarf event show that supersonic turbulence and large-scale shear suppress star formation across the bulk of the stripped gas \citep{Shin_2020, Lee_2021}, with efficient star formation occurring only in dense, shock-compressed clumps, the same conditions invoked to explain the overluminous GC populations of DF2 and DF4 \citep{Silk_2019, Lee_2021}. Adopting a global star formation efficiency of $\epsilon_{\rm SF} \sim 10$--$30\%$, consistent with the range explored in the simulations, the total gas mass required to produce the observed trail is $M_{\rm gas,\,total} \sim 2$--$8 \times 10^9\,{\rm M_\odot}$. Assuming the two progenitors contributed comparable gas masses, each would have carried a gas reservoir of $M_{\rm gas,\,prog} \sim 1$--$4 \times 10^9\,{\rm M_\odot}$, consistent with the values assumed in \citet{van_Dokkum_2022} and \citet{Lee_2024}.

To put this in perspective, we can now independently estimate the stellar mass of LEDA at the time of the collision using the same surviving stellar mass fraction prescription as above. LEDA formed the bulk of its stars at $z\sim 6$ ($t \approx 13$\,Gyr ago; Section~\ref{sec:results}). Consequently, its stellar population was approximately 5\,Gyr old at the collision epoch ($z \sim 1$) compared to 13\,Gyr old today. Evaluating the DSPS parametrization, we find the surviving stellar mass fraction was $f_{\rm surv}(5\,{\rm Gyr}) \approx 0.62$ at $z \sim 1$, compared to $f_{\rm surv}(13\,{\rm Gyr}) \approx 0.58$ at $z = 0$. Correcting for this difference yields a negligible increase in the stellar mass at the collision epoch of only $\approx 6\%$ ($\Delta \log M_\star \approx +0.03$\,dex). This result is almost insensitive to the assumed star formation history: even for a continuous SFH extending over the full 5\,Gyr prior to quenching, the implied correction rises marginally, remaining well within the measurement uncertainty on $M_\star$ ($\pm 0.06$\,dex). We therefore adopt $\log M_{\star}(z \sim 1) = 8.01 \pm 0.06$.

Comparing the gas mass available to each progenitor to LEDA's stellar mass at the collision epoch gives a gas-to-stellar mass ratio of $M_{\rm gas,\,prog}/M_\star(z \sim 1) \sim 10$--$40$. Defining the gas fraction as $f_{\rm gas} \equiv M_{\rm gas} / (M_{\rm gas} + M_\star)$, we obtain $f_{\rm gas} \sim 0.90$--$0.98$ for LEDA in a progenitor scenario. This extremely high gas fraction indicates that, if LEDA is indeed the progenitor, it was overwhelmingly gas-dominated at the time of the collision.
These extreme gas fractions are, however, physically well-motivated for a galaxy of this mass and redshift. \citet{Bradford_2015} measure a median atomic gas fraction of $f_{\rm gas} = 0.81 \pm 0.13$ for isolated galaxies in the local Universe with stellar masses between $10^{7}$ and $10^{9.5}\,M_\odot$, a range that encompasses LEDA. \citet{Zhang_2021} show that gas fractions increase systematically with redshift at fixed stellar mass, and, similarly, demonstrate that $f_{\rm gas}$ increases as $M_\star$ decreases. Taken together, these trends indicate that the local-Universe value of $f_{\rm gas} \approx 0.81$ should be regarded as a conservative lower limit for a dwarf galaxy at $z \sim 1$, making the inferred gas fraction of $f_{\rm gas} \sim 0.9$ consistent with expectations. We note that LEDA may not be well approximated as an isolated galaxy, given its possible location within the NGC\,1052 group; for non-isolated systems the scatter in $f_{\rm gas}$ presented by \citet{Bradford_2015} is larger, but the values reported ($f_{\rm gas} = 0.82 \pm 0.16$) are still consistent with the median reported for isolated systems. 

\subsubsection{The Dark Matter Halo}
\label{sec:disc_dm_halo}

An additional constraint on whether LEDA is an interloper or a progenitor remnant comes from its DM halo. If LEDA is the surviving progenitor of one of the two colliding dwarfs, it experienced an abnormal quenching event that halted its stellar mass growth at $z \sim 1$, while simultaneously stripping $\sim 90\% $ of its baryonic mass quasi-instantaneously. As a consequence, LEDA's observed stellar mass would fall below the value expected for a halo of its mass under normal evolutionary conditions, making it more DM-dominated than a typical dwarf galaxy at the same stellar mass. 
To quantitatively assess this scenario, we consider two classes of 
model: a stars-only (DM-free) case, in which the total mass is 
contributed entirely by the stellar component, and a full DM halo 
case assuming a standard Navarro--Frenk--White \citep[NFW;][]{NFW_1997} 
density profile. For the latter, we map the stellar mass to a halo 
mass $M_h$ via the redshift-dependent SMHM relation of 
\citet{Moster_2013}, compute the virial radius $r_{200}$ from the 
critical density $\rho_{\rm crit}(z)$, and derive the NFW 
concentration from the mass- and redshift-dependent relation of 
\citet{Dutton_2014}. The enclosed DM mass within $r_{1/2}$ is then 
computed from the NFW profile and converted to a predicted 
line-of-sight velocity dispersion via the \citet{Wolf_2010} mass 
estimator. To propagate all sources of uncertainty, we perform a 
Monte Carlo simulation with $N = 10^5$ realisations, drawing for 
each: (i) the stellar mass within its $1\sigma$ uncertainty; (ii) 
the intrinsic scatter in the SMHM relation 
($\sigma_{\rm SMHM} = 0.25$\,dex); (iii) the log-normal scatter in 
the concentration--mass relation ($\sigma_{\log c} = 0.11$\,dex; 
\citealt{Dutton_2014}); (iv) the effective radius within its 
measurement uncertainty; and (v) a 10\% systematic uncertainty on 
the Wolf mass estimator \citep{Wolf_2010}.

We apply both classes of model to LEDA and to the three trail DMDGs 
(DF2, DF4, and DF9) at $z \sim 0$. For the progenitor hypothesis, 
we additionally evaluate LEDA at the collision epoch ($z \sim 1$), 
using the stellar mass corrected to that epoch 
($\log M_\star = 8.01 \pm 0.06$; Section~\ref{sec:disc_gas}) and 
evaluating all cosmological quantities at $z \sim 1$.
The results are shown in Fig.~\ref{fig:forest_plot}, which is divided into two panels. The left panel shows the predictions for LEDA under both the interloper ($z \sim 0$) and progenitor ($z \sim 1$) scenarios, with the measured velocity dispersion indicated as an orange dashed line with yellow shading denoting the $1\sigma$ uncertainty. The right panel shows the equivalent analysis for the three trail DMDGs, serving as a comparison baseline, with their observed velocity dispersions similarly shown as dashed lines with shaded $1\sigma$ bands. For each galaxy and each scenario, both the stars-only (triangles) and NFW halo (squares) predictions are shown as points, with thick and thin bars indicating the 68\% and 95\% confidence intervals of the Monte Carlo distributions, respectively. From the right panel, it is clear that the present-day stellar mass alone is sufficient to account for the internal kinematics of DF2, DF4, and DF9, while a full NFW DM halo would systematically overpredict their velocity dispersions, inconsistent with the observations. This stands in stark contrast to the left panel, where the stars-only model entirely fails to reproduce LEDA's measured $\sigma$. The contrast is particularly compelling given that all four galaxies lie at similar distances within the NGC\,1052 group. This comparison therefore simultaneously confirms the DM deficiency of DF2, DF4, and DF9 and the DM dominance of LEDA.

Adopting a NFW halo, both the $z \sim 0$ and $z \sim 1$ predictions are consistent with LEDA's measured velocity dispersion at the $1\sigma$ level. The $z \sim 1$ prediction yields a marginally higher expected dispersion than the $z \sim 0$ case, a consequence of LEDA's larger inferred stellar mass at that epoch mapping onto a more massive DM halo via the SMHM relation; at first glance this makes the $z \sim 1$ scenario appear the less favourable match, but neither scenario can be distinguished from the other with the present data. 

Beyond the Monte Carlo scatter, the NFW-based predictions carry 
substantial additional uncertainty driven by the intrinsic scatter 
in the concentration--mass relation, which is particularly pronounced 
at higher redshifts \citep{Dutton_2014}. A further caveat concerns 
the assumed density profile: while a NFW cusp is a standard and 
reasonable choice for the interloper scenario, where LEDA would 
be a passively evolving dwarf unlikely to have developed a DM core 
through stellar feedback, it is a less justified assumption for the 
progenitor scenario. In that case, the true post-collision density 
profile is unknown: no dedicated simulations of the bullet-dwarf 
collision currently exist that track the response of the DM halo 
to the quasi-instantaneous removal of $\sim 90\%$ of the baryonic 
mass, and an impulsive perturbation of this magnitude could in 
principle have significantly disturbed the inner density structure and altered the observed velocity dispersion \citep[e.g.,][]{Read_2005, Li_2022}. We therefore adopt NFW as a 
working assumption, while acknowledging that the predicted $\sigma$ 
for the progenitor scenario carries an additional, unquantified 
systematic uncertainty from this unknown profile shape and from the possibility that the velocity dispersion itself has been altered. The  $z \sim 1$ prediction further assumes that LEDA's halo was frozen at the collision epoch; in practice, tidal stripping by NGC\,1052 over the  subsequent $\sim 8$\,Gyr may have gradually  reduced the enclosed DM mass relative to this prediction. The comparison between the scenarios in 
Fig.~\ref{fig:forest_plot} should therefore be regarded as indicative 
rather than a strict test.

\subsubsection{Spatial Position and Trail Membership}
\label{sec:disc_spatial}

A final constraint comes from the spatial distribution of the trail itself, as visualised in Fig.~\ref{fig:fov}. \citet{Lee_2024} predict that at least one surviving progenitor remnant should lie at one of the two extreme ends of the trail, with the caveat that the second remnant's position is not strictly constrained, as it could have been ejected at an angle or captured by the gravitational potential of NGC\,1052. Being spatially coincident with the trail therefore strengthens, though does not guarantee, the progenitor case.

Before examining LEDA's position within the trail, we note that 
its physical association with the trail is not obvious a priori: as a more 
compact and luminous galaxy, LEDA could in principle be a foreground 
or background interloper unrelated to the NGC\,1052 group entirely. 
The distance measurements of \citet{Tang_2026} place LEDA at 
essentially the same distance as DF4, reinforcing the stark kinematic 
contrast with the trail DMDGs seen in Fig.~\ref{fig:forest_plot}: 
because all galaxies share approximately the same distance, the difference in 
velocity dispersion between LEDA and DF2, DF4, and DF9 provides stronger evidence 
that the DMDGs are genuinely DM-deficient.

Examining the trail ordered by right ascension, as shown in Fig.~\ref{fig:fov}, LEDA is clearly a spatial member of the structure, but it is not the westernmost galaxy: it sits $4.7\arcmin$ east of DF7, which occupies the extreme western end and has itself been flagged as a potential interloper or progenitor \citep{Keim_2025}. As the outermost member, DF7 would be the natural candidate for the surviving progenitor remnant predicted by \citet{Lee_2024}. However, it fails this hypothesis at the first test: its mass-weighted age of $t = 6.73^{+3.74}_{-3.30}$\,Gyr \citep{Tang_2025} implies that its stellar population was still forming after the gas should have been stripped, inconsistent with a system quenched by the collision. Furthermore, its low stellar mass ($\log M_\star/M_\odot = 7.02$) implies that, as a progenitor, it would have required an even more extreme gas-dominated baryonic budget than LEDA at $z \sim 1$, further straining the hypothesis. Finally, DF7 is misaligned with the primary spatial axis defined by the remaining trail members, suggesting it may not be a genuine member of the trail at all.

The interpretation of DF7 therefore has direct consequences for LEDA's status. If future kinematic observations confirm that DF7 is a DMDG formed directly in the collision, then the trail extends further west than currently assumed, making the interloper interpretation of LEDA the more statistically straightforward one. Conversely, if DF7 is DM dominated, as would be expected for a normal dwarf galaxy unrelated to the collision, it is more naturally identified as a line-of-sight interloper, in which case LEDA becomes the outermost western member of the trail, consistent with the positional prediction of \citet{Lee_2024} and lending further circumstantial support to the progenitor hypothesis.

\subsubsection{Closing: Distinguishing the Scenarios}
\label{sec:disc_closing}

Taken together, LEDA's properties remain consistent with the progenitor remnant hypothesis across every line of evidence we have examined. At the same time, the chance-projection scenario is the statistically simpler explanation, since a chance-projection requires no specific physical properties or causal link beyond a random spatial coincidence with the DMDGs. Nevertheless, with the data currently in hand, we cannot definitively distinguish between the two. Fundamentally, the progenitor hypothesis is predicated on the physical reality of the trail itself; if the apparent trail is merely a visual alignment of the DMDGs, the progenitor scenario is inherently ruled out.

Several observational and theoretical advances would help resolve 
this question. The most important theoretical step is targeted 
$N$-body and hydrodynamical simulations of the bullet-dwarf 
collision that track the post-encounter fate of the putative progenitors' 
collisionless components. As discussed in Section~\ref{sec:disc_dm_halo}, 
the response of the DM halo to the quasi-instantaneous removal of 
$\sim 90\%$ of the baryonic mass is currently unconstrained, and 
the NFW profile adopted here is a working assumption rather than 
a physically motivated prediction. 
Such simulations would clarify which observations are best suited to distinguish between the two scenarios, while also providing the quantitative predictions needed to meaningfully test the progenitor hypothesis: the expected DM density profile, morphology, spatial distribution, and geometry of the surviving remnant.

Guided by those predictions, deeper and spatially resolved observations of LEDA and the other galaxies along the proposed trail would provide the observational tests needed to assess the viability of the bullet-dwarf scenario and LEDA's true nature. These tests include tighter constraints on the stellar population of LEDA, where the detection of a significant component younger than $\sim 8$\,Gyr would strongly favour the 
chance-alignment scenario, and direct mapping of the DM density 
profile beyond $1\,R_\mathrm{e}$. A wider spectroscopic survey 
of the NGC\,1052 group would also place LEDA's properties in 
statistical context: if LEDA proves to be the only group member 
combining a DM halo with an old, quenched, compact 
stellar population, this would strengthen the case for its physical 
association with the proposed trail, whereas the discovery of other group 
galaxies sharing these properties would weaken the significance 
of any spatial coincidence.

\section{Conclusions}
\label{sec:conclusion}

We have presented the first measurement of the stellar velocity dispersion of LEDA\,4014647, a compact, old galaxy spatially aligned with the putative NGC\,1052 bullet dwarf collision trail. Using integral field spectroscopy obtained with KCWI on the Keck\,II telescope, we extracted an integrated spectrum within $1\,R_\mathrm{e}$ and derived the kinematics and stellar populations via full-spectrum fitting with \textsc{pPXF}. Our main conclusions are as follows.

\begin{enumerate}

    \item We measure a stellar velocity dispersion of \mbox{$\sigma = 15.8 \pm 2.0$\,km\,s$^{-1}$} and find no evidence for ordered rotation across a range of position angles, indicating that LEDA is a purely dispersion-supported system.

    \item Using the \citet{Wolf_2010} mass estimator, we derive a dynamical mass within the 3D half-light radius of \mbox{$M_\mathrm{dyn}(<r_{1/2}) = (1.5 \pm 0.4) \times 10^8\,M_\odot$}, yielding a dynamical-to-stellar mass ratio of \mbox{$M_\mathrm{dyn}/M_\star = 3 \pm 1$} and a DM fraction of \mbox{$f_\mathrm{DM} = 0.7 \pm 0.1$ within $r_{1/2}$}. LEDA is therefore a DM-dominated system.

    \item The stellar populations of LEDA are old ($t = 13.2^{+0.4}_{-3.0}$\,Gyr) 
and metal-poor ($[\mathrm{M/H}] = -1.22 \pm 0.32$\,dex),, consistent with the independent photometric analysis of \citet{Tang_2025}. These properties predate the putative bullet dwarf collision epoch of $\sim 8$\,Gyr ago, ruling out formation from the shock-compressed gas hypothesised to have produced the trail DMDGs.

    \item The DM content of LEDA immediately excludes it from the population of DMDGs. However, its complete set of observed properties — DM-dominated, old, metal-poor, compact, dispersion-supported, and spatially coincident with the western end of the trail — is consistent with the expected characteristics of a surviving progenitor remnant: one of the two original DM-rich dwarfs that collided $\sim 8$\,Gyr ago to produce the trail. Alternatively, this exact set of properties is also entirely consistent with LEDA being an ordinary dwarf member of the NGC\,1052 group, with its spatial alignment being purely a chance-projection.

\end{enumerate}

Ultimately, the physical reality of the bullet-dwarf collision and its associated trail remains a hypothesis awaiting definitive confirmation.
Resolving whether LEDA is a chance-projection or the surviving collisionless remnant of such an event will require progress on both the observational and theoretical fronts. 
On the observational side, deeper spectroscopy of LEDA is needed 
to rule out any stellar population younger than $\sim 8$\,Gyr and 
place tighter constraints on its star formation history; spatially 
resolved kinematics extending beyond $1\,R_\mathrm{e}$ would map the DM 
density profile and test whether the enclosed halo mass is consistent 
with the $z \sim 1$ SMHM prediction; and a systematic spectroscopic 
survey of the remaining trail members, including a definitive 
kinematic characterisation of DF7, would establish their stellar 
populations, DM content, and group membership on a uniform footing. 
On the theoretical side, targeted $N$-body and hydrodynamical 
simulations of the bullet-dwarf collision, specifically tracking 
the post-encounter fate of the progenitors' collisionless components, 
are needed to provide quantitative predictions for the expected 
kinematics, morphology, and final position of a surviving remnant, that are 
required to ultimately test the viability of the bullet-dwarf scenario.

\section*{Acknowledgements}

We thank M. L. Buzzo, M. Monaci, B. van Heumen, A. Ferré-Mateu, and the AGATE team for stimulating discussions and valuable comments on this work.
D.V., D.F., and J.P.B. acknowledge financial support from the Australian Research Council (ARC) through Discovery Project grant DP250101673.
A.L. acknowledges support through a Swinburne University Postgraduate Research Award (SUPRA). AJR was supported by National Science Foundation grant AST-2308390. This work was supported by a NASA Keck PI Data Award, administered by the NASA Exoplanet Science Institute.
The data presented herein were obtained at the W.\,M.\ Keck Observatory, which is operated as a scientific partnership among the California Institute of Technology, the University of California, and the National Aeronautics and Space Administration; the Observatory was made possible by the generous financial support of the W.\,M.\ Keck Foundation.
The authors wish to recognise and acknowledge the very significant cultural role and reverence that the summit of Maunakea has always had within the indigenous Hawaiian community. We are most fortunate to have the opportunity to conduct observations from this mountain.
This research made use of the following software and services:
\textsc{Astropy}, a community-developed core Python package for astronomy \citep{astropy, astropy2018};
\textsc{MontagePy} \citep{Berriman_2017}, funded by the National Science Foundation under Grant Number ACI-1440620, and previously funded by the National Aeronautics and Space Administration's Earth Science Technology Office under Cooperative Agreement Number NCC5-626 between NASA and the California Institute of Technology;
\textsc{corner.py} \citep{FM2013};
\textsc{matplotlib} \citep{Hunter:2007};
\textsc{NumPy} \citep{harris2020array};
\textsc{SciPy} \citep{2020SciPy-NMeth};
and NASA's Astrophysics Data System Bibliographic Services.

\section*{Data Availability}

The KCWI integral field spectroscopy data presented in this article are available via the Keck Observatory Archive (KOA)\footnote{\url{https://www2.keck.hawaii.edu/koa/public/koa.php}} 18 months after the observations were taken.



\bibliographystyle{mnras}
\bibliography{biblio} 

@article{Shen_2021,
   title={A Tip of the Red Giant Branch Distance of 22.1 ± 1.2 Mpc to the Dark Matter Deficient Galaxy NGC 1052–DF2 from 40 Orbits of Hubble Space Telescope Imaging},
   volume={914},
   ISSN={2041-8213},
   url={http://dx.doi.org/10.3847/2041-8213/ac0335},
   DOI={10.3847/2041-8213/ac0335},
   number={1},
   journal={The Astrophysical Journal Letters},
   publisher={American Astronomical Society},
   author={Shen, Zili and Danieli, Shany and van Dokkum, Pieter and Abraham, Roberto and Brodie, Jean P. and Conroy, Charlie and Dolphin, Andrew E. and Romanowsky, Aaron J. and Diederik Kruijssen, J. M. and Dutta Chowdhury, Dhruba},
   year={2021},
   month=jun, pages={L12} }

@article{van_Dokkum_2019,
   title={A Second Galaxy Missing Dark Matter in the NGC 1052 Group},
   volume={874},
   ISSN={2041-8213},
   url={http://dx.doi.org/10.3847/2041-8213/ab0d92},
   DOI={10.3847/2041-8213/ab0d92},
   number={1},
   journal={The Astrophysical Journal Letters},
   publisher={American Astronomical Society},
   author={van Dokkum, Pieter and Danieli, Shany and Abraham, Roberto and Conroy, Charlie and Romanowsky, Aaron J.},
   year={2019},
   month=mar, pages={L5} }

@article{Danieli_2019,
   title={Still Missing Dark Matter: KCWI High-resolution Stellar Kinematics of NGC1052-DF2},
   volume={874},
   ISSN={2041-8213},
   url={http://dx.doi.org/10.3847/2041-8213/ab0e8c},
   DOI={10.3847/2041-8213/ab0e8c},
   number={2},
   journal={The Astrophysical Journal Letters},
   publisher={American Astronomical Society},
   author={Danieli, Shany and van Dokkum, Pieter and Conroy, Charlie and Abraham, Roberto and Romanowsky, Aaron J.},
   year={2019},
   month=mar, pages={L12} }

@ARTICLE{Danieli_2020,
       author = {{Danieli}, Shany and {van Dokkum}, Pieter and {Abraham}, Roberto and {Conroy}, Charlie and {Dolphin}, Andrew E. and {Romanowsky}, Aaron J.},
        title = "{A Tip of the Red Giant Branch Distance to the Dark Matter Deficient Galaxy NGC 1052-DF4 from Deep Hubble Space Telescope Data}",
      journal = {\apjl},
         year = 2020,
        month = may,
       volume = {895},
       number = {1},
          eid = {L4},
        pages = {L4},
          doi = {10.3847/2041-8213/ab8dc4},
archivePrefix = {arXiv},
       eprint = {1910.07529},
 primaryClass = {astro-ph.GA},
       adsurl = {https://ui.adsabs.harvard.edu/abs/2020ApJ...895L...4D}
}

@article{van_Dokkum_2018,
   title={A galaxy lacking dark matter},
   volume={555},
   ISSN={1476-4687},
   url={http://dx.doi.org/10.1038/nature25767},
   DOI={10.1038/nature25767},
   number={7698},
   journal={Nature},
   publisher={Springer Science and Business Media LLC},
   author={van Dokkum, Pieter and Danieli, Shany and Cohen, Yotam and Merritt, Allison and Romanowsky, Aaron J. and Abraham, Roberto and Brodie, Jean and Conroy, Charlie and Lokhorst, Deborah and Mowla, Lamiya and O’Sullivan, Ewan and Zhang, Jielai},
   year={2018},
   month=mar, pages={629–632} }

@article{Simon_2019,
   title={The Faintest Dwarf Galaxies},
   volume={57},
   ISSN={1545-4282},
   url={http://dx.doi.org/10.1146/annurev-astro-091918-104453},
   DOI={10.1146/annurev-astro-091918-104453},
   number={1},
   journal={Annual Review of Astronomy and Astrophysics},
   publisher={Annual Reviews},
   author={Simon, Joshua D.},
   year={2019},
   month=aug, pages={375–415} }

@article{Bullock_2017,
   title={Small-Scale Challenges to the ΛCDM Paradigm},
   volume={55},
   ISSN={1545-4282},
   url={http://dx.doi.org/10.1146/annurev-astro-091916-055313},
   DOI={10.1146/annurev-astro-091916-055313},
   number={1},
   journal={Annual Review of Astronomy and Astrophysics},
   publisher={Annual Reviews},
   author={Bullock, James S. and Boylan-Kolchin, Michael},
   year={2017},
   month=aug, pages={343–387} }

@article{Boylan_Kolchin_2011,
   title={Too big to fail? The puzzling darkness of massive Milky Way subhaloes},
   volume={415},
   ISSN={1745-3925},
   url={http://dx.doi.org/10.1111/j.1745-3933.2011.01074.x},
   DOI={10.1111/j.1745-3933.2011.01074.x},
   number={1},
   journal={Monthly Notices of the Royal Astronomical Society: Letters},
   publisher={Oxford University Press (OUP)},
   author={Boylan-Kolchin, Michael and Bullock, James S. and Kaplinghat, Manoj},
   year={2011},
   month=jul, pages={L40–L44} }

@BOOK{Mo_2010,
       author = {{Mo}, Houjun and {van den Bosch}, Frank C. and {White}, Simon},
        title = "{Galaxy Formation and Evolution}",
         year = 2010,
    publisher = "Cambridge University Press)",
          doi = {10.1017/CBO9780511807244},
       adsurl = {https://ui.adsabs.harvard.edu/abs/2010gfe..book.....M}
}

@article{Ogiya_2018,
   title={Tidal stripping as a possible origin of the ultra diffuse galaxy lacking dark matter},
   volume={480},
   ISSN={1745-3933},
   url={http://dx.doi.org/10.1093/mnrasl/sly138},
   DOI={10.1093/mnrasl/sly138},
   number={1},
   journal={Monthly Notices of the Royal Astronomical Society: Letters},
   publisher={Oxford University Press (OUP)},
   author={Ogiya, Go},
   year={2018},
   month=July, pages={L106–L110} }

@article{Moreno_2022,
   title={Galaxies lacking dark matter produced by close encounters in a cosmological simulation},
   volume={6},
   ISSN={2397-3366},
   url={http://dx.doi.org/10.1038/s41550-021-01598-4},
   DOI={10.1038/s41550-021-01598-4},
   number={4},
   journal={Nature Astronomy},
   publisher={Springer Science and Business Media LLC},
   author={Moreno, Jorge and Danieli, Shany and Bullock, James S. and Feldmann, Robert and Hopkins, Philip F. and Çatmabacak, Onur and Gurvich, Alexander and Lazar, Alexandres and Klein, Courtney and Hummels, Cameron B. and Hafen, Zachary and Mercado, Francisco J. and Yu, Sijie and Jiang, Fangzhou and Wheeler, Coral and Wetzel, Andrew and Anglés-Alcázar, Daniel and Boylan-Kolchin, Michael and Quataert, Eliot and Faucher-Giguère, Claude-André and Kereš, Dušan},
   year={2022},
   month=Feb, pages={496–502} }

@ARTICLE{Jackson_2021,
       author = {{Jackson}, R.~A. and {Kaviraj}, S. and {Martin}, G. and {Devriendt}, J.~E.~G. and {Slyz}, A. and {Silk}, J. and {Dubois}, Y. and {Yi}, S.~K. and {Pichon}, C. and {Volonteri}, M. and {Choi}, H. and {Kimm}, T. and {Kraljic}, K. and {Peirani}, S.},
        title = "{Dark matter-deficient dwarf galaxies form via tidal stripping of dark matter in interactions with massive companions}",
      journal = {\mnras},
         year = 2021,
        month = apr,
       volume = {502},
       number = {2},
        pages = {1785-1796},
          doi = {10.1093/mnras/stab093},
archivePrefix = {arXiv},
       eprint = {2010.02219},
 primaryClass = {astro-ph.GA},
       adsurl = {https://ui.adsabs.harvard.edu/abs/2021MNRAS.502.1785J}
}

@article{Ogiya_2021,
   title={On the tidal formation of dark matter-deficient galaxies},
   volume={510},
   ISSN={1365-2966},
   url={http://dx.doi.org/10.1093/mnras/stab3658},
   DOI={10.1093/mnras/stab3658},
   number={2},
   journal={Monthly Notices of the Royal Astronomical Society},
   publisher={Oxford University Press (OUP)},
   author={Ogiya, Go and van den Bosch, Frank C and Burkert, Andreas},
   year={2021},
   month=Dec, pages={2724–2739} }

@article{Clowe_2006,
   title={A Direct Empirical Proof of the Existence of Dark Matter},
   volume={648},
   ISSN={1538-4357},
   url={http://dx.doi.org/10.1086/508162},
   DOI={10.1086/508162},
   number={2},
   journal={The Astrophysical Journal},
   publisher={American Astronomical Society},
   author={Clowe, Douglas and Bradač, Maruša and Gonzalez, Anthony H. and Markevitch, Maxim and Randall, Scott W. and Jones, Christine and Zaritsky, Dennis},
   year={2006},
   month=Aug, pages={L109–L113} }

@ARTICLE{Shin_2020,
       author = {{Shin}, Eun-jin and {Jung}, Minyong and {Kwon}, Goojin and {Kim}, Ji-hoon and {Lee}, Joohyun and {Jo}, Yongseok and {Oh}, Boon Kiat},
        title = "{Dark Matter Deficient Galaxies Produced via High-velocity Galaxy Collisions in High-resolution Numerical Simulations}",
      journal = {\apj},
         year = 2020,
        month = aug,
       volume = {899},
       number = {1},
          eid = {25},
        pages = {25},
          doi = {10.3847/1538-4357/aba434},
archivePrefix = {arXiv},
       eprint = {2007.09889},
 primaryClass = {astro-ph.GA},
       adsurl = {https://ui.adsabs.harvard.edu/abs/2020ApJ...899...25S}
}

@ARTICLE{Silk_2019,
       author = {{Silk}, Joseph},
        title = "{Ultra-diffuse galaxies without dark matter}",
      journal = {\mnras},
         year = 2019,
        month = sep,
       volume = {488},
       number = {1},
        pages = {L24-L28},
          doi = {10.1093/mnrasl/slz090},
archivePrefix = {arXiv},
       eprint = {1905.13235},
 primaryClass = {astro-ph.GA},
       adsurl = {https://ui.adsabs.harvard.edu/abs/2019MNRAS.488L..24S}
}

@ARTICLE{Lee_2021,
       author = {{Lee}, Joohyun and {Shin}, Eun-jin and {Kim}, Ji-hoon},
        title = "{Dark Matter Deficient Galaxies and Their Member Star Clusters Form Simultaneously during High-velocity Galaxy Collisions in 1.25 pc Resolution Simulations}",
      journal = {\apjl},
         year = 2021,
        month = aug,
       volume = {917},
       number = {2},
          eid = {L15},
        pages = {L15},
          doi = {10.3847/2041-8213/ac16e0},
archivePrefix = {arXiv},
       eprint = {2108.01102},
 primaryClass = {astro-ph.GA},
       adsurl = {https://ui.adsabs.harvard.edu/abs/2021ApJ...917L..15L}
}

@article{van_Dokkum_2022,
   title={A trail of dark-matter-free galaxies from a bullet-dwarf collision},
   volume={605},
   ISSN={1476-4687},
   url={http://dx.doi.org/10.1038/s41586-022-04665-6},
   DOI={10.1038/s41586-022-04665-6},
   number={7910},
   journal={Nature},
   publisher={Springer Science and Business Media LLC},
   author={van Dokkum, Pieter and Shen, Zili and Keim, Michael A. and Trujillo-Gomez, Sebastian and Danieli, Shany and Dutta Chowdhury, Dhruba and Abraham, Roberto and Conroy, Charlie and Kruijssen, J. M. Diederik and Nagai, Daisuke and Romanowsky, Aaron},
   year={2022},
   month=May, pages={435–439} }

@article{Lee_2024,
   title={Multiple Beads on a String: Dark-matter-deficient Galaxy Formation in a Mini-Bullet Satellite–Satellite Galaxy Collision},
   volume={966},
   ISSN={1538-4357},
   url={http://dx.doi.org/10.3847/1538-4357/ad2932},
   DOI={10.3847/1538-4357/ad2932},
   number={1},
   journal={The Astrophysical Journal},
   publisher={American Astronomical Society},
   author={Lee, Joohyun and Shin, Eun-jin and Kim, Ji-hoon and Shapiro, Paul R. and Chung, Eunwoo},
   year={2024},
   month=Apr, pages={72} }

@article{Keim_2025,
doi = {10.3847/1538-4357/addfd4},
url = {https://doi.org/10.3847/1538-4357/addfd4},
year = {2025},
month = {jul},
publisher = {The American Astronomical Society},
volume = {988},
number = {2},
pages = {165},
author = {Keim, Michael A. and van Dokkum, Pieter and Shen, Zili and Souchereau, Harrison and Pasha, Imad and Danieli, Shany and Abraham, Roberto and Romanowsky, Aaron J. and Tang, Yimeng},
title = {Kinematic Confirmation of a Remarkable Linear Trail of Galaxies in the NGC 1052 Field, Consistent with Formation in a High-speed Bullet Dwarf Collision},
journal = {The Astrophysical Journal}
}

@article{Fensch_2019,
   title={The ultra-diffuse galaxy NGC 1052-DF2 with MUSE: II. The population of DF2: stars, clusters, and planetary nebulae},
   volume={625},
   ISSN={1432-0746},
   url={http://dx.doi.org/10.1051/0004-6361/201834911},
   DOI={10.1051/0004-6361/201834911},
   journal={Astronomy & Astrophysics},
   publisher={EDP Sciences},
   author={Fensch, Jérémy and van der Burg, Remco F. J. and Jeřábková, Tereza and Emsellem, Eric and Zanella, Anita and Agnello, Adriano and Hilker, Michael and Müller, Oliver and Rejkuba, Marina and Duc, Pierre-Alain and Durrell, Patrick and Habas, Rebecca and Lim, Sungsoon and Marleau, Francine R. and Peng, Eric W. and Sánchez Janssen, Rubén},
   year={2019},
   month=May, pages={A77} }

@ARTICLE{van_Dokkum_2022b,
       author = {{van Dokkum}, Pieter and {Shen}, Zili and {Romanowsky}, Aaron J. and {Abraham}, Roberto and {Conroy}, Charlie and {Danieli}, Shany and {Chowdhury}, Dhruba Dutta and {Keim}, Michael A. and {Kruijssen}, J.~M. Diederik and {Leja}, Joel and {Trujillo-Gomez}, Sebastian},
        title = "{Monochromatic Globular Clusters as a Critical Test of Formation Models for the Dark Matter-deficient Galaxies NGC 1052-DF2 and NGC 1052-DF4}",
      journal = {\apjl},
         year = 2022,
        month = nov,
       volume = {940},
       number = {1},
          eid = {L9},
        pages = {L9},
          doi = {10.3847/2041-8213/ac94d6},
archivePrefix = {arXiv},
       eprint = {2207.07129},
 primaryClass = {astro-ph.GA},
       adsurl = {https://ui.adsabs.harvard.edu/abs/2022ApJ...940L...9V}
}

@ARTICLE{Lara_2019,
       author = {{Ruiz-Lara}, T. and {Trujillo}, I. and {Beasley}, M.~A. and {Falc{\'o}n-Barroso}, J. and {Vazdekis}, A. and {Filho}, M. and {Monelli}, M. and {Rom{\'a}n}, J. and {S{\'a}nchez Almeida}, J.},
        title = "{Stellar content, planetary nebulae, and globular clusters of [KKS2000]04 (NGC 1052-DF2)}",
      journal = {\mnras},
         year = 2019,
        month = jul,
       volume = {486},
       number = {4},
        pages = {5670-5678},
          doi = {10.1093/mnras/stz1237},
archivePrefix = {arXiv},
       eprint = {1903.09163},
 primaryClass = {astro-ph.GA},
       adsurl = {https://ui.adsabs.harvard.edu/abs/2019MNRAS.486.5670R}
}

@article{Buzzo_2022,
    author = {Buzzo, Maria Luisa and Forbes, Duncan A and Brodie, Jean P and Romanowsky, Aaron J and Cluver, Michelle E and Jarrett, Thomas H and Laine, Seppo and Couch, Warrick J and Gannon, Jonah S and Ferré-Mateu, Anna and Okabe, Nobuhiro},
    title = {The stellar populations of quiescent ultra-diffuse galaxies from optical to mid-infrared spectral energy distribution fitting},
    journal = {Monthly Notices of the Royal Astronomical Society},
    volume = {517},
    number = {2},
    pages = {2231-2250},
    year = {2022},
    month = {12},
    issn = {0035-8711},
    doi = {10.1093/mnras/stac2442},
    url = {https://doi.org/10.1093/mnras/stac2442},
    eprint = {https://academic.oup.com/mnras/article-pdf/517/2/2231/46563510/stac2442.pdf},
}

@article{Tang_2025,
doi = {10.3847/1538-4357/ad8cd0},
url = {https://doi.org/10.3847/1538-4357/ad8cd0},
year = {2025},
month = {jan},
publisher = {The American Astronomical Society},
volume = {978},
number = {1},
pages = {21},
author = {Tang, Yimeng and Romanowsky, Aaron J. and van Dokkum, Pieter G. and Jarrett, T. H. and Bundy, Kevin A. and Buzzo, Maria Luisa and Danieli, Shany and Gannon, Jonah S. and Keim, Michael A. and Laine, Seppo and Shen, Zili},
title = {Testing the Bullet Dwarf Collision Scenario in the NGC 1052 Group through Morphologies and Stellar Populations},
journal = {The Astrophysical Journal}
}

@article{Tang_2025_224,
   title={An Unexplained Origin for the Unusual Globular Cluster System in the Ultradiffuse Galaxy FCC 224},
   volume={982},
   ISSN={1538-4357},
   url={http://dx.doi.org/10.3847/1538-4357/adae11},
   DOI={10.3847/1538-4357/adae11},
   number={1},
   journal={The Astrophysical Journal},
   publisher={American Astronomical Society},
   author={Tang, Yimeng and Romanowsky, Aaron J. and Gannon, Jonah S. and Janssens, Steven R. and Brodie, Jean P. and Bundy, Kevin A. and Buzzo, Maria Luisa and Cabrera, Enrique A. and Danieli, Shany and Ferré-Mateu, Anna and Forbes, Duncan A. and van Dokkum, Pieter G.},
   year={2025},
   month=Mar, pages={1} }

@article{Romanowsky_2024,
doi = {10.3847/2515-5172/ad7112},
url = {https://doi.org/10.3847/2515-5172/ad7112},
year = {2024},
month = {aug},
publisher = {The American Astronomical Society},
volume = {8},
number = {8},
pages = {202},
author = {Romanowsky, Aaron J. and Cabrera, Enrique and Janssens, Steven R.},
title = {A Candidate Dark Matter Deficient Dwarf Galaxy in the Fornax Cluster Identified through Overluminous Star Clusters},
journal = {Research Notes of the AAS}
}

@article{Buzzo_2025,
   title={A new class of dark matter-free dwarf galaxies?: I. Clues from FCC 224, NGC 1052-DF2, and NGC 1052-DF4},
   volume={695},
   ISSN={1432-0746},
   url={http://dx.doi.org/10.1051/0004-6361/202453522},
   DOI={10.1051/0004-6361/202453522},
   journal={Astronomy & Astrophysics},
   publisher={EDP Sciences},
   author={Buzzo, Maria Luisa and Forbes, Duncan A. and Romanowsky, Aaron J. and Haacke, Lydia and Gannon, Jonah S. and Tang, Yimeng and Hilker, Michael and Ferré-Mateu, Anna and Janssens, Steven R. and Brodie, Jean P. and Valenzuela, Lucas M.},
   year={2025},
   month=Mar, pages={A124} }

@article{Buzzo_2026,
doi = {10.3847/1538-4357/ae7349},
url = {https://doi.org/10.3847/1538-4357/ae7349},
year = {2026},
month = {jun},
publisher = {The American Astronomical Society},
volume = {1005},
number = {1},
pages = {15},
author = {Buzzo, Maria Luísa and van Dokkum, Pieter and Hilker, Michael and Forbes, Duncan A. and Romanowsky, Aaron J. and Tang, Yimeng},
title = {Dark-matter-deficient Twins: FCC 224 and FCC 240 as Possible Analogs of NGC 1052-DF2 and DF4},
journal = {The Astrophysical Journal},
}

@misc{Keim_2026,
      title={A Third Galaxy Missing Dark Matter along a Trail of Galaxies in the NGC 1052 Field}, 
      author={Michael A. Keim and Pieter van Dokkum and Zili Shen and Shany Danieli and Imad Pasha},
      year={2026},
      eprint={2603.15860},
      archivePrefix={arXiv},
      primaryClass={astro-ph.GA},
      url={https://arxiv.org/abs/2603.15860}, 
}

@article{Morrissey_2018,
doi = {10.3847/1538-4357/aad597},
url = {https://doi.org/10.3847/1538-4357/aad597},
year = {2018},
month = {sep},
publisher = {The American Astronomical Society},
volume = {864},
number = {1},
pages = {93},
author = {Morrissey, Patrick and Matuszewski, Matuesz and Martin, D. Christopher and Neill, James D. and Epps, Harland and Fucik, Jason and Weber, Bob and Darvish, Behnam and Adkins, Sean and Allen, Steve and Bartos, Randy and Belicki, Justin and Cabak, Jerry and Callahan, Shawn and Cowley, Dave and Crabill, Marty and Deich, Willian and Delecroix, Alex and Doppman, Greg and Hilyard, David and James, Ean and Kaye, Steve and Kokorowski, Michael and Kwok, Shui and Lanclos, Kyle and Milner, Steve and Moore, Anna and O’Sullivan, Donal and Parihar, Prachi and Park, Sam and Phillips, Andrew and Rizzi, Luca and Rockosi, Constance and Rodriguez, Hector and Salaun, Yves and Seaman, Kirk and Sheikh, David and Weiss, Jason and Zarzaca, Ray},
title = {The Keck Cosmic Web Imager Integral Field Spectrograph},
journal = {The Astrophysical Journal}
}

@ARTICLE{Cappellari_2004,
       author = {{Cappellari}, Michele and {Emsellem}, Eric},
        title = "{Parametric Recovery of Line-of-Sight Velocity Distributions from Absorption-Line Spectra of Galaxies via Penalized Likelihood}",
      journal = {\pasp},
         year = 2004,
        month = feb,
       volume = {116},
       number = {816},
        pages = {138-147},
          doi = {10.1086/381875},
archivePrefix = {arXiv},
       eprint = {astro-ph/0312201},
 primaryClass = {astro-ph},
       adsurl = {https://ui.adsabs.harvard.edu/abs/2004PASP..116..138C}
}

@ARTICLE{Cappellari_2017,
       author = {{Cappellari}, Michele},
        title = "{Improving the full spectrum fitting method: accurate convolution with Gauss-Hermite functions}",
      journal = {\mnras},
         year = 2017,
        month = apr,
       volume = {466},
       number = {1},
        pages = {798-811},
          doi = {10.1093/mnras/stw3020},
archivePrefix = {arXiv},
       eprint = {1607.08538},
 primaryClass = {astro-ph.GA},
       adsurl = {https://ui.adsabs.harvard.edu/abs/2017MNRAS.466..798C}
}

@ARTICLE{Cappellari_2023,
       author = {{Cappellari}, Michele},
        title = "{Full spectrum fitting with photometry in PPXF: stellar population versus dynamical masses, non-parametric star formation history and metallicity for 3200 LEGA-C galaxies at redshift z {\ensuremath{\approx}} 0.8}",
      journal = {\mnras},
         year = 2023,
        month = dec,
       volume = {526},
       number = {3},
        pages = {3273-3300},
          doi = {10.1093/mnras/stad2597},
archivePrefix = {arXiv},
       eprint = {2208.14974},
 primaryClass = {astro-ph.GA},
       adsurl = {https://ui.adsabs.harvard.edu/abs/2023MNRAS.526.3273C}
}

@article{Coelho_2014,
   title={A new library of theoretical stellar spectra with scaled-solar and α-enhanced mixtures},
   volume={440},
   ISSN={1365-2966},
   url={http://dx.doi.org/10.1093/mnras/stu365},
   DOI={10.1093/mnras/stu365},
   number={2},
   journal={Monthly Notices of the Royal Astronomical Society},
   publisher={Oxford University Press (OUP)},
   author={Coelho, P. R. T.},
   year={2014},
   month=Apr, pages={1027–1043} 
}

@ARTICLE{Vazdekis_2015,
       author = {{Vazdekis}, A. and {Coelho}, P. and {Cassisi}, S. and {Ricciardelli}, E. and {Falc{\'o}n-Barroso}, J. and {S{\'a}nchez-Bl{\'a}zquez}, P. and {La Barbera}, F. and {Beasley}, M.~A. and {Pietrinferni}, A.},
        title = "{Evolutionary stellar population synthesis with MILES - II. Scaled-solar and {\ensuremath{\alpha}}-enhanced models}",
      journal = {\mnras},
         year = 2015,
        month = may,
       volume = {449},
       number = {2},
        pages = {1177-1214},
          doi = {10.1093/mnras/stv151},
archivePrefix = {arXiv},
       eprint = {1504.08032},
 primaryClass = {astro-ph.GA},
       adsurl = {https://ui.adsabs.harvard.edu/abs/2015MNRAS.449.1177V}
}

@ARTICLE{Kroupa_2001,
       author = {{Kroupa}, Pavel},
        title = "{On the variation of the initial mass function}",
      journal = {\mnras},
         year = 2001,
        month = apr,
       volume = {322},
       number = {2},
        pages = {231-246},
          doi = {10.1046/j.1365-8711.2001.04022.x},
archivePrefix = {arXiv},
       eprint = {astro-ph/0009005},
 primaryClass = {astro-ph},
       adsurl = {https://ui.adsabs.harvard.edu/abs/2001MNRAS.322..231K}
}

@ARTICLE{Hidalgo_2018,
       author = {{Hidalgo}, Sebastian L. and {Pietrinferni}, Adriano and {Cassisi}, Santi and {Salaris}, Maurizio and {Mucciarelli}, Alessio and {Savino}, Alessandro and {Aparicio}, Antonio and {Silva Aguirre}, Victor and {Verma}, Kuldeep},
        title = "{The Updated BaSTI Stellar Evolution Models and Isochrones. I. Solar-scaled Calculations}",
      journal = {\apj},
         year = 2018,
        month = apr,
       volume = {856},
       number = {2},
          eid = {125},
        pages = {125},
          doi = {10.3847/1538-4357/aab158},
archivePrefix = {arXiv},
       eprint = {1802.07319},
 primaryClass = {astro-ph.GA},
       adsurl = {https://ui.adsabs.harvard.edu/abs/2018ApJ...856..125H}
}

@Article{Wolf_2010,
  author    = {Wolf, Joe and Martinez, Gregory D. and Bullock, James S. and Kaplinghat, Manoj and Geha, Marla and Muñoz, Ricardo R. and Simon, Joshua D. and Avedo, Frank F.},
  journal   = {\mnras},
  title     = {Accurate masses for dispersion-supported galaxies},
  year      = {2010},
  issn      = {1365-2966},
  month     = {May},
  pages     = {1220-1237},
  doi       = {10.1111/j.1365-2966.2010.16753.x},
  publisher = {Oxford University Press (OUP)},
  url       = {http://dx.doi.org/10.1111/j.1365-2966.2010.16753.x},
}

@article{Behroozi_2013,
   title={THE AVERAGE STAR FORMATION HISTORIES OF GALAXIES IN DARK MATTER HALOS FROM<i>z</i>= 0-8},
   volume={770},
   ISSN={1538-4357},
   url={http://dx.doi.org/10.1088/0004-637X/770/1/57},
   DOI={10.1088/0004-637x/770/1/57},
   number={1},
   journal={The Astrophysical Journal},
   publisher={American Astronomical Society},
   author={Behroozi, Peter S. and Wechsler, Risa H. and Conroy, Charlie},
   year={2013},
   month=May, pages={57} }

@ARTICLE{Koda_2015,
       author = {{Koda}, Jin and {Yagi}, Masafumi and {Yamanoi}, Hitomi and {Komiyama}, Yutaka},
        title = "{Approximately a Thousand Ultra-diffuse Galaxies in the Coma Cluster}",
      journal = {\apjl},
         year = 2015,
        month = jul,
       volume = {807},
       number = {1},
          eid = {L2},
        pages = {L2},
          doi = {10.1088/2041-8205/807/1/L2},
archivePrefix = {arXiv},
       eprint = {1506.01712},
 primaryClass = {astro-ph.GA},
       adsurl = {https://ui.adsabs.harvard.edu/abs/2015ApJ...807L...2K}
}

@ARTICLE{van_der_Burg_2016,
       author = {{van der Burg}, Remco F.~J. and {Muzzin}, Adam and {Hoekstra}, Henk},
        title = "{The abundance and spatial distribution of ultra-diffuse galaxies in nearby galaxy clusters}",
      journal = {\aap},
         year = 2016,
        month = may,
       volume = {590},
          eid = {A20},
        pages = {A20},
          doi = {10.1051/0004-6361/201628222},
archivePrefix = {arXiv},
       eprint = {1602.00002},
 primaryClass = {astro-ph.GA},
       adsurl = {https://ui.adsabs.harvard.edu/abs/2016A&A...590A..20V}
}

@ARTICLE{Roman_2017,
       author = {{Rom{\'a}n}, Javier and {Trujillo}, Ignacio},
        title = "{Ultra-diffuse galaxies outside clusters: clues to their formation and evolution}",
      journal = {\mnras},
         year = 2017,
        month = jul,
       volume = {468},
       number = {4},
        pages = {4039-4047},
          doi = {10.1093/mnras/stx694},
archivePrefix = {arXiv},
       eprint = {1610.08980},
 primaryClass = {astro-ph.GA},
       adsurl = {https://ui.adsabs.harvard.edu/abs/2017MNRAS.468.4039R}
}

@ARTICLE{Wasserman_2018,
       author = {{Wasserman}, Asher and {Romanowsky}, Aaron J. and {Brodie}, Jean and {van Dokkum}, Pieter and {Conroy}, Charlie and {Abraham}, Roberto and {Cohen}, Yotam and {Danieli}, Shany},
        title = "{A Deficit of Dark Matter from Jeans Modeling of the Ultra-diffuse Galaxy NGC 1052-DF2}",
      journal = {\apjl},
         year = 2018,
        month = aug,
       volume = {863},
       number = {2},
          eid = {L15},
        pages = {L15},
          doi = {10.3847/2041-8213/aad779},
archivePrefix = {arXiv},
       eprint = {1807.07069},
 primaryClass = {astro-ph.GA},
       adsurl = {https://ui.adsabs.harvard.edu/abs/2018ApJ...863L..15W}
}

@ARTICLE{Emsellem_2019,
       author = {{Emsellem}, Eric and {van der Burg}, Remco F.~J. and {Fensch}, J{\'e}r{\'e}my and {Je{\v{r}}{\'a}bkov{\'a}}, Tereza and {Zanella}, Anita and {Agnello}, Adriano and {Hilker}, Michael and {M{\"u}ller}, Oliver and {Rejkuba}, Marina and {Duc}, Pierre-Alain and {Durrell}, Patrick and {Habas}, Rebecca and {Lelli}, Federico and {Lim}, Sungsoon and {Marleau}, Francine R. and {Peng}, Eric and {S{\'a}nchez-Janssen}, Rub{\'e}n},
        title = "{The ultra-diffuse galaxy NGC 1052-DF2 with MUSE. I. Kinematics of the stellar body}",
      journal = {\aap},
         year = 2019,
        month = may,
       volume = {625},
          eid = {A76},
        pages = {A76},
          doi = {10.1051/0004-6361/201834909},
archivePrefix = {arXiv},
       eprint = {1812.07345},
 primaryClass = {astro-ph.GA},
       adsurl = {https://ui.adsabs.harvard.edu/abs/2019A&A...625A..76E}
}

@ARTICLE{van_Dokkum_2015,
       author = {{van Dokkum}, Pieter G. and {Abraham}, Roberto and {Merritt}, Allison and {Zhang}, Jielai and {Geha}, Marla and {Conroy}, Charlie},
        title = "{Forty-seven Milky Way-sized, Extremely Diffuse Galaxies in the Coma Cluster}",
      journal = {\apjl},
         year = 2015,
        month = jan,
       volume = {798},
       number = {2},
          eid = {L45},
        pages = {L45},
          doi = {10.1088/2041-8205/798/2/L45},
archivePrefix = {arXiv},
       eprint = {1410.8141},
 primaryClass = {astro-ph.GA},
       adsurl = {https://ui.adsabs.harvard.edu/abs/2015ApJ...798L..45V}
}

@article{Tanoglidis_2021,
   title={Shadows in the Dark: Low-surface-brightness Galaxies Discovered in the Dark Energy Survey},
   volume={252},
   ISSN={1538-4365},
   url={http://dx.doi.org/10.3847/1538-4365/abca89},
   DOI={10.3847/1538-4365/abca89},
   number={2},
   journal={The Astrophysical Journal Supplement Series},
   publisher={American Astronomical Society},
   author={Tanoglidis, D. and Drlica-Wagner, A. and Wei, K. and Li, T. S. and Sánchez, J. and Zhang, Y. and Peter, A. H. G. and Feldmeier-Krause, A. and Prat, J. and Casey, K. and Palmese, A. and Sánchez, C. and DeRose, J. and Conselice, C. and Gagnon, L. and Abbott, T. M. C. and Aguena, M. and Allam, S. and Avila, S. and Bechtol, K. and Bertin, E. and Bhargava, S. and Brooks, D. and Burke, D. L. and Rosell, A. Carnero and Kind, M. Carrasco and Carretero, J. and Chang, C. and Costanzi, M. and Costa, L. N. da and Vicente, J. De and Desai, S. and Diehl, H. T. and Doel, P. and Eifler, T. F. and Everett, S. and Evrard, A. E. and Flaugher, B. and Frieman, J. and García-Bellido, J. and Gerdes, D. W. and Gruendl, R. A. and Gschwend, J. and Gutierrez, G. and Hartley, W. G. and Hollowood, D. L. and Huterer, D. and James, D. J. and Krause, E. and Kuehn, K. and Kuropatkin, N. and Maia, M. A. G. and March, M. and Marshall, J. L. and Menanteau, F. and Miquel, R. and Ogando, R. L. C. and Paz-Chinchón, F. and Romer, A. K. and Roodman, A. and Sanchez, E. and Scarpine, V. and Serrano, S. and Sevilla-Noarbe, I. and Smith, M. and Suchyta, E. and Tarle, G. and Thomas, D. and Tucker, D. L. and Walker, A. R. and },
   year={2021},
   month=Jan, pages={18} }

@article{Zaritsky_2022,
   title={Systematically Measuring Ultra-diffuse Galaxies (SMUDGes). III. The Southern SMUDGes Catalog},
   volume={261},
   ISSN={1538-4365},
   url={http://dx.doi.org/10.3847/1538-4365/ac6ceb},
   DOI={10.3847/1538-4365/ac6ceb},
   number={2},
   journal={The Astrophysical Journal Supplement Series},
   publisher={American Astronomical Society},
   author={Zaritsky, Dennis and Donnerstein, Richard and Karunakaran, Ananthan and Barbosa, C. E. and Dey, Arjun and Kadowaki, Jennifer and Spekkens, Kristine and Zhang, Huanian},
   year={2022},
   month=July, pages={11} }

@article{Davidson_2008,
title = {The wild bootstrap, tamed at last},
journal = {Journal of Econometrics},
volume = {146},
number = {1},
pages = {162-169},
year = {2008},
issn = {0304-4076},
doi = {https://doi.org/10.1016/j.jeconom.2008.08.003},
url = {https://www.sciencedirect.com/science/article/pii/S0304407608000833},
author = {Russell Davidson and Emmanuel Flachaire}
}

@ARTICLE{Gannon_2026,
       author = {{Gannon}, Jonah S. and {Ferr{\'e}-Mateu}, Anna and {Forbes}, Duncan A.},
        title = "{The Dawes review 14: A decade of ultra-diffuse galaxies}",
      journal = {\pasa},
         year = 2026,
        month = mar,
       volume = {43},
          eid = {e031},
        pages = {e031},
          doi = {10.1017/pasa.2026.10169},
archivePrefix = {arXiv},
       eprint = {2602.21875},
 primaryClass = {astro-ph.GA},
       adsurl = {https://ui.adsabs.harvard.edu/abs/2026PASA...43...31G}
}

@ARTICLE{Shen_2023,
       author = {Shen, Zili and van Dokkum, Pieter and Danieli, Shany},
        title = "{Confirmation of an Anomalously Low Dark Matter Content for the Galaxy NGC 1052-DF4 from Deep, High-resolution Continuum Spectroscopy}",
      journal = {\apj},
         year = 2023,
        month = nov,
       volume = {957},
       number = {1},
          eid = {6},
        pages = {6},
          doi = {10.3847/1538-4357/acfa70},
archivePrefix = {arXiv},
       eprint = {2309.08592},
 primaryClass = {astro-ph.GA},
       adsurl = {https://ui.adsabs.harvard.edu/abs/2023ApJ...957....6S}
}

@article{Vaz_2026,
       author = {{Vaz}, Daniel and {Brinchmann}, Jarle and {Kamann}, Sebastian and {Saracino}, Sara and {Cunha}, P.~A.~C. and {J{\'u}lio}, Mariana P.},
        title = "{The MUSE-Faint survey: V. The binary fraction of Leo T}",
      journal = {\aap},
         year = 2026,
        month = mar,
       volume = {708},
          eid = {A101},
        pages = {A101},
          doi = {10.1051/0004-6361/202555752},
archivePrefix = {arXiv},
       eprint = {2603.02643},
 primaryClass = {astro-ph.GA},
       adsurl = {https://ui.adsabs.harvard.edu/abs/2026A&A...708A.101V}
}

@ARTICLE{Gannon_2020,
       author = {{Gannon}, Jonah S. and {Forbes}, Duncan A. and {Romanowsky}, Aaron J. and {Ferr{\'e}-Mateu}, Anna and {Couch}, Warrick J. and {Brodie}, Jean P.},
        title = "{On the stellar kinematics and mass of the Virgo ultradiffuse galaxy VCC 1287}",
      journal = {\mnras},
         year = 2020,
        month = jul,
       volume = {495},
       number = {3},
        pages = {2582-2598},
          doi = {10.1093/mnras/staa1282},
archivePrefix = {arXiv},
       eprint = {2005.03041},
 primaryClass = {astro-ph.GA},
       adsurl = {https://ui.adsabs.harvard.edu/abs/2020MNRAS.495.2582G}
}

@ARTICLE{Levitskiy_2025,
       author = {{Levitskiy}, Arsen and {Forbes}, Duncan A. and {Gannon}, Jonah S. and {Ferr{\'e}-Mateu}, Anna and {Romanowsky}, Aaron J. and {Brodie}, Jean P. and {Couch}, Warrick J. and {Haacke}, Lydia},
        title = "{A comprehensive look at PUDG-R21: stellar population and kinematics of a globular cluster-rich ultra-diffuse galaxy in the Perseus Cluster}",
      journal = {\mnras},
         year = 2025,
        month = aug,
       volume = {541},
       number = {3},
        pages = {2761-2772},
          doi = {10.1093/mnras/staf1140},
archivePrefix = {arXiv},
       eprint = {2507.05679},
 primaryClass = {astro-ph.GA},
       adsurl = {https://ui.adsabs.harvard.edu/abs/2025MNRAS.541.2761L}
}

@ARTICLE{Berriman_2017,
       author = {{Berriman}, G. Bruce and {Good}, J.~C.},
        title = "{The Application of the Montage Image Mosaic Engine To The Visualization Of Astronomical Images}",
      journal = {\pasp},
         year = 2017,
        month = may,
       volume = {129},
       number = {975},
        pages = {058006},
          doi = {10.1088/1538-3873/aa5456},
archivePrefix = {arXiv},
       eprint = {1702.02593},
 primaryClass = {astro-ph.IM},
       adsurl = {https://ui.adsabs.harvard.edu/abs/2017PASP..129e8006B}
}

@ARTICLE{Dey_2019,
       author = {{Dey}, Arjun and {Schlegel}, David J. and {Lang}, Dustin and {Blum}, Robert and {Burleigh}, Kaylan and {Fan}, Xiaohui and {Findlay}, Joseph R. and {Finkbeiner}, Doug and {Herrera}, David and {Juneau}, St{\'e}phanie and {Landriau}, Martin and {Levi}, Michael and {McGreer}, Ian and {Meisner}, Aaron and {Myers}, Adam D. and {Moustakas}, John and {Nugent}, Peter and {Patej}, Anna and {Schlafly}, Edward F. and {Walker}, Alistair R. and {Valdes}, Francisco and {Weaver}, Benjamin A. and {Y{\`e}che}, Christophe and {Zou}, Hu and {Zhou}, Xu and {Abareshi}, Behzad and {Abbott}, T.~M.~C. and {Abolfathi}, Bela and {Aguilera}, C. and {Alam}, Shadab and {Allen}, Lori and {Alvarez}, A. and {Annis}, James and {Ansarinejad}, Behzad and {Aubert}, Marie and {Beechert}, Jacqueline and {Bell}, Eric F. and {BenZvi}, Segev Y. and {Beutler}, Florian and {Bielby}, Richard M. and {Bolton}, Adam S. and {Brice{\~n}o}, C{\'e}sar and {Buckley-Geer}, Elizabeth J. and {Butler}, Karen and {Calamida}, Annalisa and {Carlberg}, Raymond G. and {Carter}, Paul and {Casas}, Ricard and {Castander}, Francisco J. and {Choi}, Yumi and {Comparat}, Johan and {Cukanovaite}, Elena and {Delubac}, Timoth{\'e}e and {DeVries}, Kaitlin and {Dey}, Sharmila and {Dhungana}, Govinda and {Dickinson}, Mark and {Ding}, Zhejie and {Donaldson}, John B. and {Duan}, Yutong and {Duckworth}, Christopher J. and {Eftekharzadeh}, Sarah and {Eisenstein}, Daniel J. and {Etourneau}, Thomas and {Fagrelius}, Parker A. and {Farihi}, Jay and {Fitzpatrick}, Mike and {Font-Ribera}, Andreu and {Fulmer}, Leah and {G{\"a}nsicke}, Boris T. and {Gaztanaga}, Enrique and {George}, Koshy and {Gerdes}, David W. and {Gontcho}, Satya Gontcho A. and {Gorgoni}, Claudio and {Green}, Gregory and {Guy}, Julien and {Harmer}, Diane and {Hernandez}, M. and {Honscheid}, Klaus and {Huang}, Lijuan Wendy and {James}, David J. and {Jannuzi}, Buell T. and {Jiang}, Linhua and {Joyce}, Richard and {Karcher}, Armin and {Karkar}, Sonia and {Kehoe}, Robert and {Kneib}, Jean-Paul and {Kueter-Young}, Andrea and {Lan}, Ting-Wen and {Lauer}, Tod R. and {Le Guillou}, Laurent and {Le Van Suu}, Auguste and {Lee}, Jae Hyeon and {Lesser}, Michael and {Perreault Levasseur}, Laurence and {Li}, Ting S. and {Mann}, Justin L. and {Marshall}, Robert and {Mart{\'\i}nez-V{\'a}zquez}, C.~E. and {Martini}, Paul and {du Mas des Bourboux}, H{\'e}lion and {McManus}, Sean and {Meier}, Tobias Gabriel and {M{\'e}nard}, Brice and {Metcalfe}, Nigel and {Mu{\~n}oz-Guti{\'e}rrez}, Andrea and {Najita}, Joan and {Napier}, Kevin and {Narayan}, Gautham and {Newman}, Jeffrey A. and {Nie}, Jundan and {Nord}, Brian and {Norman}, Dara J. and {Olsen}, Knut A.~G. and {Paat}, Anthony and {Palanque-Delabrouille}, Nathalie and {Peng}, Xiyan and {Poppett}, Claire L. and {Poremba}, Megan R. and {Prakash}, Abhishek and {Rabinowitz}, David and {Raichoor}, Anand and {Rezaie}, Mehdi and {Robertson}, A.~N. and {Roe}, Natalie A. and {Ross}, Ashley J. and {Ross}, Nicholas P. and {Rudnick}, Gregory and {Safonova}, Sasha and {Saha}, Abhijit and {S{\'a}nchez}, F. Javier and {Savary}, Elodie and {Schweiker}, Heidi and {Scott}, Adam and {Seo}, Hee-Jong and {Shan}, Huanyuan and {Silva}, David R. and {Slepian}, Zachary and {Soto}, Christian and {Sprayberry}, David and {Staten}, Ryan and {Stillman}, Coley M. and {Stupak}, Robert J. and {Summers}, David L. and {Sien Tie}, Suk and {Tirado}, H. and {Vargas-Maga{\~n}a}, Mariana and {Vivas}, A. Katherina and {Wechsler}, Risa H. and {Williams}, Doug and {Yang}, Jinyi and {Yang}, Qian and {Yapici}, Tolga and {Zaritsky}, Dennis and {Zenteno}, A. and {Zhang}, Kai and {Zhang}, Tianmeng and {Zhou}, Rongpu and {Zhou}, Zhimin},
        title = "{Overview of the DESI Legacy Imaging Surveys}",
      journal = {\aj},
         year = 2019,
        month = may,
       volume = {157},
       number = {5},
          eid = {168},
        pages = {168},
          doi = {10.3847/1538-3881/ab089d},
archivePrefix = {arXiv},
       eprint = {1804.08657},
 primaryClass = {astro-ph.IM},
       adsurl = {https://ui.adsabs.harvard.edu/abs/2019AJ....157..168D}
}

@ARTICLE{astropy,
       author = {{Astropy Collaboration} and {Robitaille}, Thomas P. and {Tollerud}, Erik J. and {Greenfield}, Perry and {Droettboom}, Michael and {Bray}, Erik and {Aldcroft}, Tom and {Davis}, Matt and {Ginsburg}, Adam and {Price-Whelan}, Adrian M. and {Kerzendorf}, Wolfgang E. and {Conley}, Alexander and {Crighton}, Neil and {Barbary}, Kyle and {Muna}, Demitri and {Ferguson}, Henry and {Grollier}, Fr{\'e}d{\'e}ric and {Parikh}, Madhura M. and {Nair}, Prasanth H. and {Unther}, Hans M. and {Deil}, Christoph and {Woillez}, Julien and {Conseil}, Simon and {Kramer}, Roban and {Turner}, James E.~H. and {Singer}, Leo and {Fox}, Ryan and {Weaver}, Benjamin A. and {Zabalza}, Victor and {Edwards}, Zachary I. and {Azalee Bostroem}, K. and {Burke}, D.~J. and {Casey}, Andrew R. and {Crawford}, Steven M. and {Dencheva}, Nadia and {Ely}, Justin and {Jenness}, Tim and {Labrie}, Kathleen and {Lim}, Pey Lian and {Pierfederici}, Francesco and {Pontzen}, Andrew and {Ptak}, Andy and {Refsdal}, Brian and {Servillat}, Mathieu and {Streicher}, Ole},
        title = "{Astropy: A community Python package for astronomy}",
      journal = {\aap},
         year = 2013,
        month = oct,
       volume = {558},
          eid = {A33},
        pages = {A33},
          doi = {10.1051/0004-6361/201322068},
archivePrefix = {arXiv},
       eprint = {1307.6212},
        primaryClass = {astro-ph.IM},
       adsurl = {https://ui.adsabs.harvard.edu/abs/2013A&A...558A..33A}
}

@ARTICLE{astropy2018,
       author = {{Astropy Collaboration} and {Price-Whelan}, A.~M. and {Sip{\H{o}}cz}, B.~M. and {G{\"u}nther}, H.~M. and {Lim}, P.~L. and {Crawford}, S.~M. and {Conseil}, S. and {Shupe}, D.~L. and {Craig}, M.~W. and {Dencheva}, N. and {Ginsburg}, A. and {VanderPlas}, J.~T. and {Bradley}, L.~D. and {P{\'e}rez-Su{\'a}rez}, D. and {de Val-Borro}, M. and {Aldcroft}, T.~L. and {Cruz}, K.~L. and {Robitaille}, T.~P. and {Tollerud}, E.~J. and {Ardelean}, C. and {Babej}, T. and {Bach}, Y.~P. and {Bachetti}, M. and {Bakanov}, A.~V. and {Bamford}, S.~P. and {Barentsen}, G. and {Barmby}, P. and {Baumbach}, A. and {Berry}, K.~L. and {Biscani}, F. and {Boquien}, M. and {Bostroem}, K.~A. and {Bouma}, L.~G. and {Brammer}, G.~B. and {Bray}, E.~M. and {Breytenbach}, H. and {Buddelmeijer}, H. and {Burke}, D.~J. and {Calderone}, G. and {Cano Rodr{\'\i}guez}, J.~L. and {Cara}, M. and {Cardoso}, J.~V.~M. and {Cheedella}, S. and {Copin}, Y. and {Corrales}, L. and {Crichton}, D. and {D'Avella}, D. and {Deil}, C. and {Depagne}, {\'E}. and {Dietrich}, J.~P. and {Donath}, A. and {Droettboom}, M. and {Earl}, N. and {Erben}, T. and {Fabbro}, S. and {Ferreira}, L.~A. and {Finethy}, T. and {Fox}, R.~T. and {Garrison}, L.~H. and {Gibbons}, S.~L.~J. and {Goldstein}, D.~A. and {Gommers}, R. and {Greco}, J.~P. and {Greenfield}, P. and {Groener}, A.~M. and {Grollier}, F. and {Hagen}, A. and {Hirst}, P. and {Homeier}, D. and {Horton}, A.~J. and {Hosseinzadeh}, G. and {Hu}, L. and {Hunkeler}, J.~S. and {Ivezi{\'c}}, {\v{Z}}. and {Jain}, A. and {Jenness}, T. and {Kanarek}, G. and {Kendrew}, S. and {Kern}, N.~S. and {Kerzendorf}, W.~E. and {Khvalko}, A. and {King}, J. and {Kirkby}, D. and {Kulkarni}, A.~M. and {Kumar}, A. and {Lee}, A. and {Lenz}, D. and {Littlefair}, S.~P. and {Ma}, Z. and {Macleod}, D.~M. and {Mastropietro}, M. and {McCully}, C. and {Montagnac}, S. and {Morris}, B.~M. and {Mueller}, M. and {Mumford}, S.~J. and {Muna}, D. and {Murphy}, N.~A. and {Nelson}, S. and {Nguyen}, G.~H. and {Ninan}, J.~P. and {N{\"o}the}, M. and {Ogaz}, S. and {Oh}, S. and {Parejko}, J.~K. and {Parley}, N. and {Pascual}, S. and {Patil}, R. and {Patil}, A.~A. and {Plunkett}, A.~L. and {Prochaska}, J.~X. and {Rastogi}, T. and {Reddy Janga}, V. and {Sabater}, J. and {Sakurikar}, P. and {Seifert}, M. and {Sherbert}, L.~E. and {Sherwood-Taylor}, H. and {Shih}, A.~Y. and {Sick}, J. and {Silbiger}, M.~T. and {Singanamalla}, S. and {Singer}, L.~P. and {Sladen}, P.~H. and {Sooley}, K.~A. and {Sornarajah}, S. and {Streicher}, O. and {Teuben}, P. and {Thomas}, S.~W. and {Tremblay}, G.~R. and {Turner}, J.~E.~H. and {Terr{\'o}n}, V. and {van Kerkwijk}, M.~H. and {de la Vega}, A. and {Watkins}, L.~L. and {Weaver}, B.~A. and {Whitmore}, J.~B. and {Woillez}, J. and {Zabalza}, V. and {Astropy Contributors}},
        title = "{The Astropy Project: Building an Open-science Project and Status of the v2.0 Core Package}",
      journal = {\aj},
         year = 2018,
        month = sep,
       volume = {156},
       number = {3},
          eid = {123},
        pages = {123},
          doi = {10.3847/1538-3881/aabc4f},
archivePrefix = {arXiv},
       eprint = {1801.02634},
 primaryClass = {astro-ph.IM},
       adsurl = {https://ui.adsabs.harvard.edu/abs/2018AJ....156..123A}
}

@Article{Hunter:2007,
  Author    = {Hunter, J. D.},
  Title     = {Matplotlib: A 2D graphics environment},
  Journal   = {Computing in Science \& Engineering},
  Volume    = {9},
  Number    = {3},
  Pages     = {90--95},
  publisher = {IEEE COMPUTER SOC},
  doi       = {10.1109/MCSE.2007.55},
  year      = 2007
}

@Article{         harris2020array,
 title         = {Array programming with {NumPy}},
 author        = {Charles R. Harris and K. Jarrod Millman and St{\'{e}}fan J.
                 van der Walt and Ralf Gommers and Pauli Virtanen and David
                 Cournapeau and Eric Wieser and Julian Taylor and Sebastian
                 Berg and Nathaniel J. Smith and Robert Kern and Matti Picus
                 and Stephan Hoyer and Marten H. van Kerkwijk and Matthew
                 Brett and Allan Haldane and Jaime Fern{\'{a}}ndez del
                 R{\'{i}}o and Mark Wiebe and Pearu Peterson and Pierre
                 G{\'{e}}rard-Marchant and Kevin Sheppard and Tyler Reddy and
                 Warren Weckesser and Hameer Abbasi and Christoph Gohlke and
                 Travis E. Oliphant},
 year          = {2020},
 month         = sep,
 journal       = {Nature},
 volume        = {585},
 number        = {7825},
 pages         = {357--362},
 doi           = {10.1038/s41586-020-2649-2},
 publisher     = {Springer Science and Business Media {LLC}},
 url           = {https://doi.org/10.1038/s41586-020-2649-2}
}

@ARTICLE{2020SciPy-NMeth,
  author  = {Virtanen, Pauli and Gommers, Ralf and Oliphant, Travis E. and
            Haberland, Matt and Reddy, Tyler and Cournapeau, David and
            Burovski, Evgeni and Peterson, Pearu and Weckesser, Warren and
            Bright, Jonathan and {van der Walt}, St{\'e}fan J. and
            Brett, Matthew and Wilson, Joshua and Millman, K. Jarrod and
            Mayorov, Nikolay and Nelson, Andrew R. J. and Jones, Eric and
            Kern, Robert and Larson, Eric and Carey, C J and
            Polat, {\.I}lhan and Feng, Yu and Moore, Eric W. and
            {VanderPlas}, Jake and Laxalde, Denis and Perktold, Josef and
            Cimrman, Robert and Henriksen, Ian and Quintero, E. A. and
            Harris, Charles R. and Archibald, Anne M. and
            Ribeiro, Ant{\^o}nio H. and Pedregosa, Fabian and
            {van Mulbregt}, Paul and {SciPy 1.0 Contributors}},
  title   = {{{SciPy} 1.0: Fundamental Algorithms for Scientific
            Computing in Python}},
  journal = {Nature Methods},
  year    = {2020},
  volume  = {17},
  pages   = {261--272},
  adsurl  = {https://rdcu.be/b08Wh},
  doi     = {10.1038/s41592-019-0686-2},
}

@ARTICLE{FM2013,
       author = {{Foreman-Mackey}, Daniel and {Hogg}, David W. and {Lang}, Dustin and {Goodman}, Jonathan},
        title = "{emcee: The MCMC Hammer}",
      journal = {\pasp},
         year = 2013,
        month = mar,
       volume = {125},
       number = {925},
        pages = {306},
          doi = {10.1086/670067},
archivePrefix = {arXiv},
       eprint = {1202.3665},
 primaryClass = {astro-ph.IM},
       adsurl = {https://ui.adsabs.harvard.edu/abs/2013PASP..125..306F}
}

@article{Tang_2026,
doi = {10.3847/2041-8213/ae77ef},
url = {https://doi.org/10.3847/2041-8213/ae77ef},
year = {2026},
month = {jun},
publisher = {The American Astronomical Society},
volume = {1005},
number = {1},
pages = {L16},
author = {Tang, Yimeng and Anand, Gagandeep S. and Romanowsky, Aaron J. and van Dokkum, Pieter G. and Bundy, Kevin A.},
title = {New Measurements of Distances to Galaxies in the NGC 1052 Field with the Hubble and James Webb Space Telescopes: Testing the Bullet-dwarf Origin of the Trail},
journal = {The Astrophysical Journal Letters}
}

@article{Moster_2013,
    author = {Moster, Benjamin P. and Naab, Thorsten and White, Simon D. M.},
    title = {Galactic star formation and accretion histories from matching galaxies to dark matter haloes },
    journal = {Monthly Notices of the Royal Astronomical Society},
    volume = {428},
    number = {4},
    pages = {3121-3138},
    year = {2013},
    month = {02},
    issn = {0035-8711},
    doi = {10.1093/mnras/sts261},
    url = {https://doi.org/10.1093/mnras/sts261},
    eprint = {https://academic.oup.com/mnras/article-pdf/428/4/3121/18461767/sts261.pdf},
}

@article{Dutton_2014,
    author = {Dutton, Aaron A. and Macciò, Andrea V.},
    title = {Cold dark matter haloes in the Planck era: evolution of structural parameters for Einasto and NFW profiles},
    journal = {Monthly Notices of the Royal Astronomical Society},
    volume = {441},
    number = {4},
    pages = {3359-3374},
    year = {2014},
    month = {07},
    issn = {0035-8711},
    doi = {10.1093/mnras/stu742},
    url = {https://doi.org/10.1093/mnras/stu742},
    eprint = {https://academic.oup.com/mnras/article-pdf/441/4/3359/4043254/stu742.pdf},
}

@article{Conroy_2010,
       author = {{Conroy}, Charlie and {Gunn}, James E.},
        title = "{The Propagation of Uncertainties in Stellar Population Synthesis Modeling. III. Model Calibration, Comparison, and Evaluation}",
      journal = {\apj},
         year = 2010,
        month = apr,
       volume = {712},
       number = {2},
        pages = {833-857},
          doi = {10.1088/0004-637X/712/2/833},
archivePrefix = {arXiv},
       eprint = {0911.3151},
 primaryClass = {astro-ph.CO},
       adsurl = {https://ui.adsabs.harvard.edu/abs/2010ApJ...712..833C}
}

@article{Hearin_2023,
   title={DSPS: Differentiable stellar population synthesis},
   volume={521},
   ISSN={1365-2966},
   url={http://dx.doi.org/10.1093/mnras/stad456},
   DOI={10.1093/mnras/stad456},
   number={2},
   journal={Monthly Notices of the Royal Astronomical Society},
   publisher={Oxford University Press (OUP)},
   author={Hearin, Andrew P and Chaves-Montero, Jonás and Alarcon, Alex and Becker, Matthew R and Benson, Andrew},
   year={2023},
   month=Feb, pages={1741–1756} }

@article{Bradford_2015,
   title={A STUDY IN BLUE: THE BARYON CONTENT OF ISOLATED LOW-MASS GALAXIES},
   volume={809},
   ISSN={1538-4357},
   url={http://dx.doi.org/10.1088/0004-637X/809/2/146},
   DOI={10.1088/0004-637x/809/2/146},
   number={2},
   journal={The Astrophysical Journal},
   publisher={American Astronomical Society},
   author={Bradford, Jeremy D. and Geha, Marla C. and Blanton, Michael R.},
   year={2015},
   month=Aug, pages={146} }

@ARTICLE{Zhang_2021,
       author = {{Zhang}, Wei and {Kauffmann}, Guinevere and {Wang}, Jing and {Chen}, Yanmei and {Fu}, Jian and {Wu}, Hong},
        title = "{A study of the H I gas fractions of galaxies at z {\ensuremath{\sim}} 1}",
      journal = {\aap},
         year = 2021,
        month = apr,
       volume = {648},
          eid = {A25},
        pages = {A25},
          doi = {10.1051/0004-6361/202039878},
archivePrefix = {arXiv},
       eprint = {2011.04500},
 primaryClass = {astro-ph.GA},
       adsurl = {https://ui.adsabs.harvard.edu/abs/2021A&A...648A..25Z}
}

@ARTICLE{NFW_1997,
       author = {{Navarro}, Julio F. and {Frenk}, Carlos S. and {White}, Simon D.~M.},
        title = "{A Universal Density Profile from Hierarchical Clustering}",
      journal = {\apj},
         year = 1997,
        month = dec,
       volume = {490},
       number = {2},
        pages = {493-508},
          doi = {10.1086/304888},
archivePrefix = {arXiv},
       eprint = {astro-ph/9611107},
 primaryClass = {astro-ph},
       adsurl = {https://ui.adsabs.harvard.edu/abs/1997ApJ...490..493N}
}

@article{Li_2022,
   title={The response of dark matter haloes to gas ejection: CuspCore II},
   volume={518},
   ISSN={1365-2966},
   url={http://dx.doi.org/10.1093/mnras/stac3233},
   DOI={10.1093/mnras/stac3233},
   number={4},
   journal={Monthly Notices of the Royal Astronomical Society},
   publisher={Oxford University Press (OUP)},
   author={Li, Zhaozhou and Dekel, Avishai and Mandelker, Nir and Freundlich, Jonathan and François, Thibaut L},
   year={2022},
   month=Nov, pages={5356–5375} }

@ARTICLE{Read_2005,
       author = {{Read}, J.~I. and {Gilmore}, G.},
        title = "{Mass loss from dwarf spheroidal galaxies: the origins of shallow dark matter cores and exponential surface brightness profiles}",
      journal = {\mnras},
         year = 2005,
        month = jan,
       volume = {356},
       number = {1},
        pages = {107-124},
          doi = {10.1111/j.1365-2966.2004.08424.x},
archivePrefix = {arXiv},
       eprint = {astro-ph/0409565},
 primaryClass = {astro-ph},
       adsurl = {https://ui.adsabs.harvard.edu/abs/2005MNRAS.356..107R}
}

@ARTICLE{La_Marca_2022,
       author = {{La Marca}, Antonio and {Iodice}, Enrichetta and {Cantiello}, Michele and {Forbes}, Duncan A. and {Rejkuba}, Marina and {Hilker}, Michael and {Arnaboldi}, Magda and {Greggio}, Laura and {Spiniello}, Chiara and {Mieske}, Steffen and {Venhola}, Aku and {Spavone}, Marilena and {D'Ago}, Giuseppe and {Raj}, Maria Angela and {Ragusa}, Rossella and {Mirabile}, Marco and {Rampazzo}, Roberto and {Peletier}, Reynier and {Paolillo}, Maurizio and {Challapa}, Nelvy Choque and {Schipani}, Pietro},
        title = "{Galaxy populations in the Hydra I cluster from the VEGAS survey. II. The ultra-diffuse galaxy population}",
      journal = {\aap},
         year = 2022,
        month = sep,
       volume = {665},
          eid = {A105},
        pages = {A105},
          doi = {10.1051/0004-6361/202142367},
archivePrefix = {arXiv},
       eprint = {2206.07385},
 primaryClass = {astro-ph.GA},
       adsurl = {https://ui.adsabs.harvard.edu/abs/2022A&A...665A.105L}
}

@article{Forbes_2020,
   title={Ultradiffuse galaxies in the IC 1459 group from the VEGAS survey},
   volume={494},
   ISSN={1365-2966},
   url={http://dx.doi.org/10.1093/mnras/staa1111},
   DOI={10.1093/mnras/staa1111},
   number={4},
   journal={Monthly Notices of the Royal Astronomical Society},
   publisher={Oxford University Press (OUP)},
   author={Forbes, Duncan A and Dullo, Bililign T and Gannon, Jonah and Couch, Warrick J and Iodice, Enrichetta and Spavone, Marilena and Cantiello, Michele and Schipani, Pietro},
   year={2020},
   month=Apr, pages={5293–5297} }

@article{Trujillo_2019,
   title={A distance of 13 Mpc resolves the claimed anomalies of the galaxy lacking dark matter},
   volume={486},
   ISSN={1365-2966},
   url={http://dx.doi.org/10.1093/mnras/stz771},
   DOI={10.1093/mnras/stz771},
   number={1},
   journal={Monthly Notices of the Royal Astronomical Society},
   publisher={Oxford University Press (OUP)},
   author={Trujillo, Ignacio and Beasley, Michael A and Borlaff, Alejandro and Carrasco, Eleazar R and Di Cintio, Arianna and Filho, Mercedes and Monelli, Matteo and Montes, Mireia and Román, Javier and Ruiz-Lara, Tomás and Sánchez Almeida, Jorge and Valls-Gabaud, David and Vazdekis, Alexandre},
   year={2019},
   month=Mar, pages={1192–1219} }

@article{Monelli_2019,
   title={The TRGB Distance to the Second Galaxy “Missing Dark Matter”: Evidence for Two Groups of Galaxies at 13.5 and 19 Mpc in the Line of Sight of NGC 1052},
   volume={880},
   ISSN={2041-8213},
   url={http://dx.doi.org/10.3847/2041-8213/ab2fd2},
   DOI={10.3847/2041-8213/ab2fd2},
   number={1},
   journal={The Astrophysical Journal Letters},
   publisher={American Astronomical Society},
   author={Monelli, Matteo and Trujillo, Ignacio},
   year={2019},
   month=July, pages={L11} }

@article{Beasley_2025,
	author = {{Beasley, M. A.} and {Fahrion, K.} and {Guerra Arencibia, S.} and {Gvozdenko, A.} and {Montes, M.}},
	title = {A new way to measure the distance to NGC1052-DF2⋆},
	DOI= "10.1051/0004-6361/202452446",
	url= "https://doi.org/10.1051/0004-6361/202452446",
	journal = {A&A},
	year = 2025,
	volume = 697,
	pages = "A144",
}

@article{Fahrion_2025,
   title={Revisiting the globular clusters of NGC 1052-DF2},
   volume={697},
   ISSN={1432-0746},
   url={http://dx.doi.org/10.1051/0004-6361/202452454},
   DOI={10.1051/0004-6361/202452454},
   journal={Astronomy &amp; Astrophysics},
   publisher={EDP Sciences},
   author={Fahrion, K. and Beasley, M. A. and Gvozdenko, A. and Guerra Arencibia, S. and Jerabkova, T. and Fensch, J. and Emsellem, E.},
   year={2025},
   month=May, pages={A145} }




\appendix

\section{Bootstrap posterior distributions}\label{app:bootstrap_corner}

\begin{figure}
    \centering
    \includegraphics[width=\columnwidth]{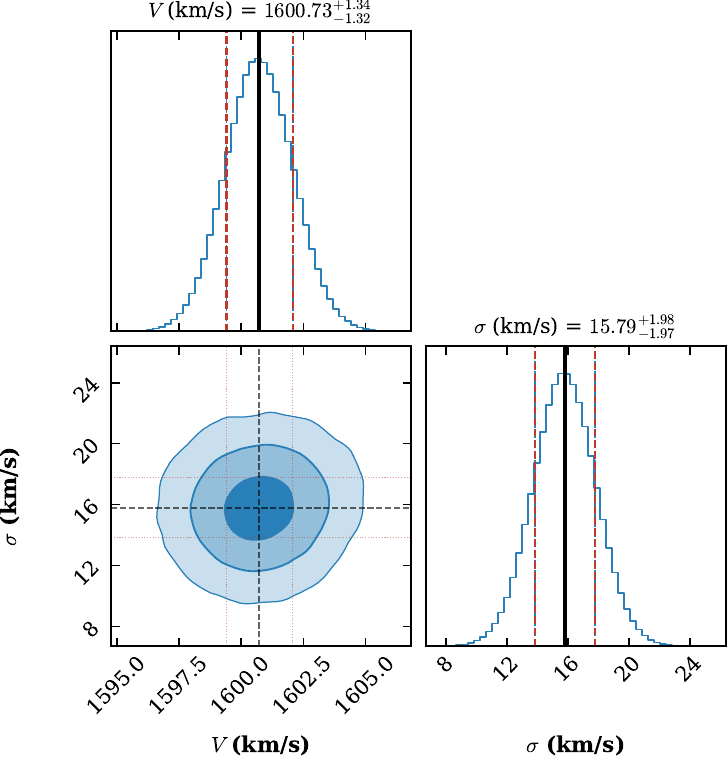}
    \caption{%
        Marginalised bootstrap distributions of the \textsc{pPXF}
        kinematic parameters $(V, \sigma)$ for LEDA, based on
        $N_\mathrm{boot} = 100{,}000$ wild bootstrap realisations
        of the $R_\mathrm{e}$-aperture spectrum.
        Diagonal panels show the marginal distributions; the
        off-diagonal panel shows the joint $(V, \sigma)$ distribution.
        Bold vertical lines indicate the bootstrap median; dashed red
        lines mark the 16th and 84th percentiles.
        Contours enclose 39.3\%, 86.5\%, and 98.9\% of the
        realisations (equivalent to 1, 2, and 3\,$\sigma$ of a
        2D Gaussian).
    }
    \label{fig:kin_corner}
\end{figure}

\begin{figure}
    \centering
    \includegraphics[width=\columnwidth]{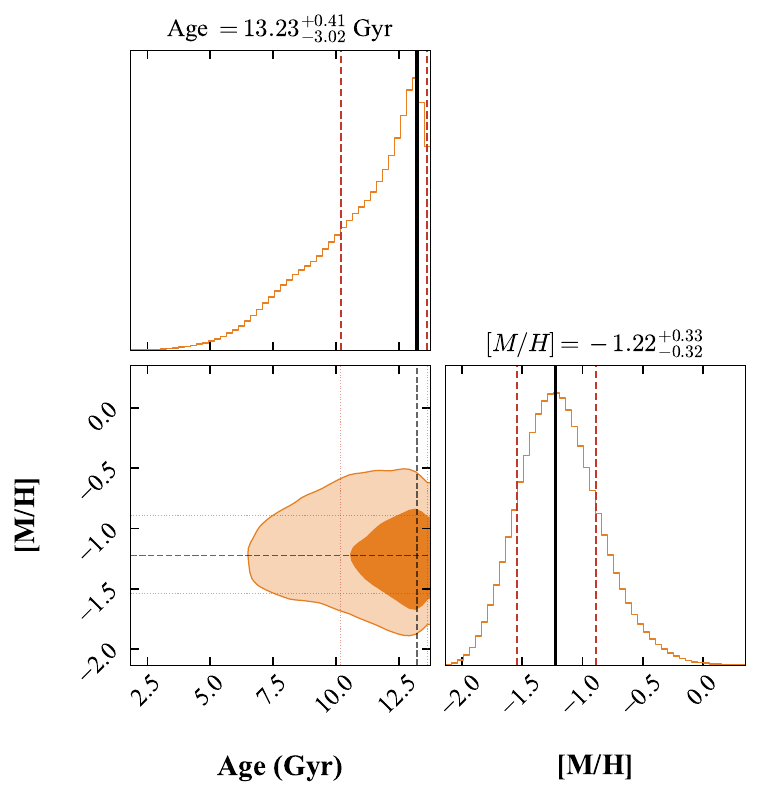}
    \caption{%
        Marginalised bootstrap distributions of the mass-weighted
        mean age and metallicity of LEDA, based on
        $N_\mathrm{boot} = 100{,}000$ wild bootstrap realisations
        of the $R_\mathrm{e}$-aperture spectrum fitted with E-MILES SSP templates.
        Diagonal panels show the marginal distributions; the
        off-diagonal panel shows the joint distribution.
        Bold vertical lines indicate the mode of each distribution,
        estimated via kernel density estimation; dashed red lines mark
        the bounds of the 68 per cent HDI.
        Contours enclose 39.3\% and 86.5\% of the realisations
        (equivalent to 1 and 2\,$\sigma$ of a 2D Gaussian).
        The broad distributions reflect the limited stellar population
        discriminating power of the available spectral range at
        SNR\,$\approx 15$ per pixel.
    }
    \label{fig:pop_corner}
\end{figure}

\begin{figure*}
    \centering
    \includegraphics[width=\textwidth]{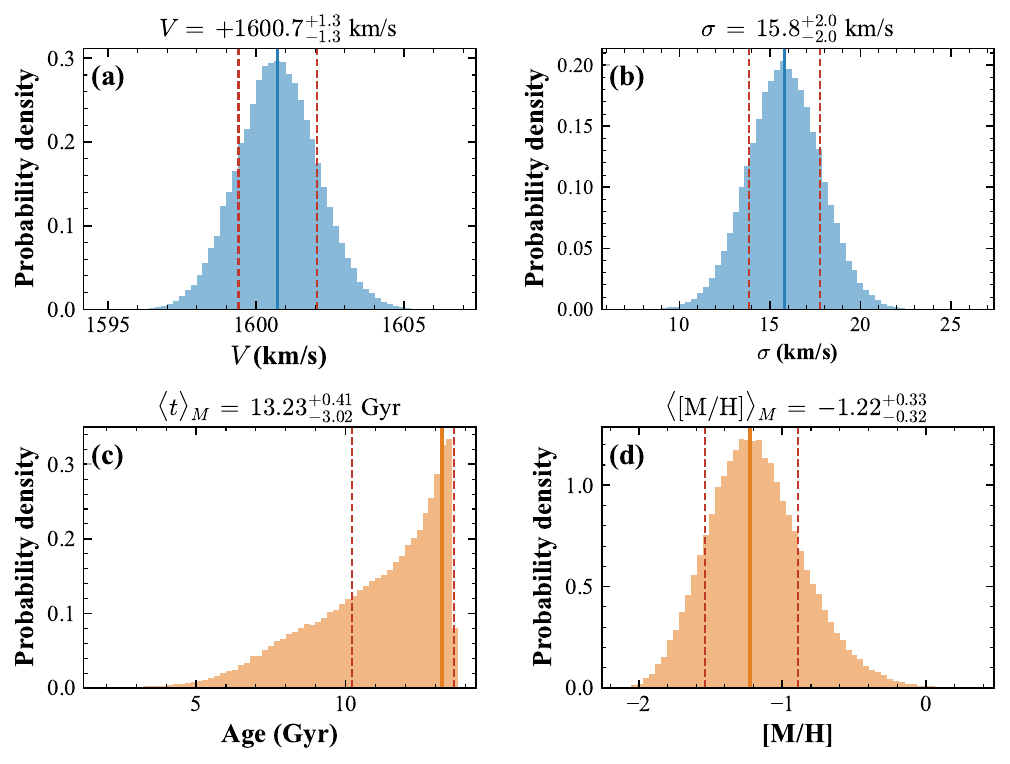}
    \caption{%
        Bootstrap probability distributions for the four \textsc{pPXF}
        output parameters of LEDA: (a) systemic velocity $V$, (b) velocity
        dispersion $\sigma$, (c) mass-weighted mean age, and (d)
        mean metallicity $[\mathrm{M/H}]$, based on
        $N_\mathrm{boot} = 100{,}000$ wild bootstrap realisations
        of the $R_\mathrm{e}$-aperture spectrum.
        Solid vertical lines indicate the mode of each distribution;
        dashed lines mark the bounds of the 68 per cent Highest Density
        Interval, quoted as subtitles above each panel.
        Blue and orange histograms show the kinematic and stellar
        population parameters, respectively.
    }
    \label{fig:summary}
\end{figure*}

We estimated uncertainties on all \textsc{pPXF} output parameters using a wild bootstrap procedure \citep[e.g.][]{Davidson_2008}. In each of the $N_\mathrm{boot} = 100{,}000$ realisations, the best-fit residuals were multiplied element-wise by an independent vector of random sign flips drawn from a Rademacher distribution, and the perturbed spectrum was re-fitted with \textsc{pPXF} using the same configuration as the original fit. This approach preserves the autocorrelation structure of the residuals and is robust to non-Gaussian noise, making it well suited to low-SNR spectra. The marginalised and joint posterior distributions for the kinematic and stellar population parameters are shown in Figs~\ref{fig:kin_corner}--\ref{fig:pop_corner}. Summary distributions for all four parameters are additionally shown in Fig.~\ref{fig:summary}.

\begin{figure}
    \centering
    \includegraphics[width=\columnwidth]{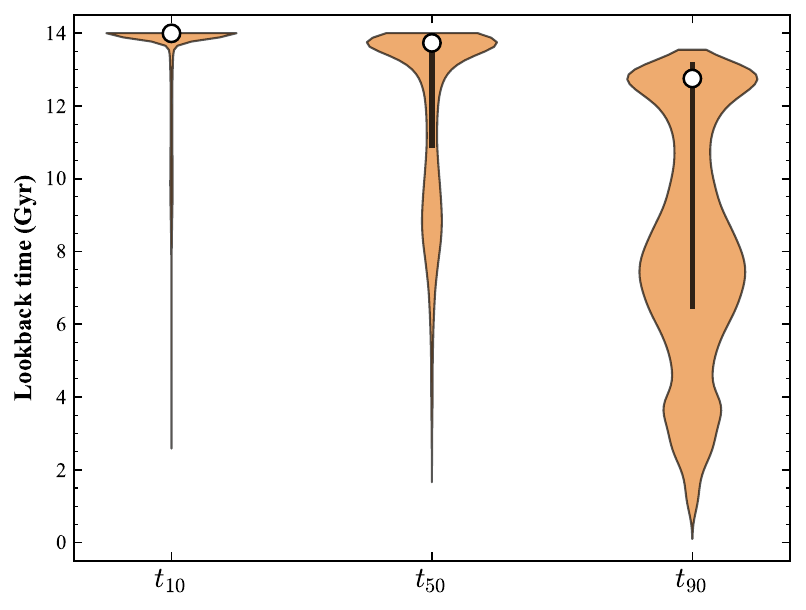}
    \caption{%
        Violin plots showing the bootstrap posterior distributions of the lookback times at which 10, 50, and 90 per cent of the stellar mass was assembled ($t_{10}$, $t_{50}$, and $t_{90}$). While the distributions are heavily skewed towards the oldest available templates (with kernel density modes at 14.0, 13.7, and 12.8\,Gyr), the tails toward younger ages reflect a subset of fits that favour a secondary star formation episode around $\sim 8$\,Gyr ago.
    }
    \label{fig:violin_plots}
\end{figure}

To further characterise the star formation history, we tracked the lookback times at which 10, 50, and 90 per cent of the stellar mass had formed ($t_{10}$, $t_{50}$, and $t_{90}$) for each individual bootstrap realisation. The resulting distributions are shown as violin plots in Fig.~\ref{fig:violin_plots}. The pronounced peaks highlight a strong preference for an ancient, single-burst star formation history, with the modes of the distributions situated at 14.0, 13.7, and 12.8\,Gyr, respectively. However, the extended tails towards younger ages indicate that in a subset of the realisations the fit prefers an extended SFH. This is a direct consequence of the age--metallicity degeneracy in our low SNR spectrum, though the overarching conclusion that LEDA is a predominantly old system remains robust across all iterations.


\bsp	
\label{lastpage}
\end{document}